\documentclass[preprint,nofootinbib,floats,eqsecnum,showpacs,11pt]{revtex4-1}
\usepackage{bm}
\usepackage{amsmath}
\usepackage{hyperref} 
\usepackage{color} 
\usepackage{amsfonts}
\usepackage{amssymb}
\usepackage{braket}

\begin{document}

\title{Quantum Cauchy Surfaces in Canonical Quantum Gravity}

\author{Chun-Yen Lin}
\email{chunyenlin@cc.ncu.edu.tw} \affiliation{Department of Physics, National Central University, Jhongli, Taoyuan 32001, Taiwan}

\begin{abstract}
\textbf{For a Dirac theory of quantum gravity obtained from the refined algebraic quantization procedure, we propose a quantum notion of Cauchy surfaces. In such a theory, there is a kernel projector for the quantized scalar and momentum constraints, which maps the kinematic Hilbert space $\mathbb K$ into the physical Hilbert space $\mathbb H$. Under this projection, a quantum Cauchy surface isomorphically represents $\mathbb H$ with a kinematic subspace $\mathbb V \subset\mathbb K$. The isomorphism induces the complete sets of Dirac observables in $\mathbb D$, which faithfully represent the corresponding complete sets of self-adjoint operators in $\mathbb V$. Due to the constraints, a specific subset of the observables would be ``frozen" as number operators, providing a background physical time for the rest of the observables. Therefore, a proper foliation with the quantum Cauchy surfaces may provide an observer frame describing the physical states of spacetimes in a Schr\"odinger picture, with the evolutions under a specific physical background. A simple model will be supplied as an initiative trial.
}
\end{abstract}

\maketitle

\section{Introduction}

The Schr\"odinger picture is crucial for our understanding and experimental testing of the existing quantum theories. In this picture, the dynamics is the evolution of the wave functions over the spectrum of a certain complete set of observables, happening under a physical background that provides the notion of time. As it is well known, a Schr\"odinger theory is a quantum representation of a Hamiltonian dynamics.

In quantum gravity, finding the Schr\"odinger representation poses profound challenges \cite{kuchartorre}\cite{Isham}. Canonical general relativity had succeeded in providing an initial data formulation for the globally hyperbolic spacetimes. However, since the theory has no fixed physical background, formulating it as a Hamiltonian dynamics in a physical time demands a splitting of the phase space coordinates, into two subsets of the background fields and the dynamical fields \cite{kuchar}\cite{torre1}. Particularly, the background has to yield a monotonic physical time, and the dynamical fields have to define a phase space of complete observables. Under gravity, the background and the dynamical fields must be coupled, and are constrained by one another through the scalar and momentum constraints. Therefore, the proper splitting becomes a highly nontrivial and state dependent issue 
\cite{kuchartorre}\cite{Isham}. At the quantum level, a direct Schr\"odinger representation is thus hindered for a quantum theory of general relativity. 

In addressing this problem, the existing theories of canonical quantum gravity mainly follow either one of the two approaches we now briefly describe. Throughout the paper, we will allow matter fields in a theory of our concern. To focus on the gravitational issues, we will assume a kinematic phase space which already solves all the constraints except for the scalar and momentum constraints $\{C^{\text N}_i\}$, respectively labeled by $\text N=0 $ and $\text N=1,2,3 $. The index $i$ labels the spatial modes in a compact space. 

The first type of theories follows the reduced phase space quantization \cite{reduced1}\cite{reduced2}\cite{reduced3}, and they are Schr\"odinger theories by construction. Starting from the classical level, one first identifies a physically significant subset $\mathbb D^{cl}$ of the physical states, which allows the proper splitting of the phase space coordinates mentioned above. We denote the chosen background fields as $\{\phi^{\text N}_i, P^{\text N}_i\}$, and the dynamical fields as $\{\psi^{K}_i, P^{K}_i\}$. For the full theory, the background consists of four clock fields $\{\phi^{\text N}_i\}$ which coordinatize each of the constraint orbits on the constraint surface belonging to $\mathbb D^{cl}$. These background fields then provide a set of physical spacetime coordinates to each of the spacetimes represented by the orbits.  We may then define the set of dynamical observables $\{\psi^{K}_i(\bar\phi_j^{\text N}), P^{K}_i(\bar\phi_j^{\text N})\}$ as functions in $\mathbb D^{cl}$ taking the values of the dynamical fields at where $\{\phi^{\text N}_j=\bar{\phi}^{\text N}_j\}$ in each of the spacetimes. Meanwhile, the observables $\{\phi_i^{\text N}(\bar\phi_j^{\text N})\}$ are trivialy defined, and the values of $\{P_i^{\text N}(\bar\phi_j^{\text N})\}$ are determined by solving the constraints. Remarkably, for each allowed value of $\bar\phi_j^{\text N}$ the dynamical observables $\{\psi^{K}_i(\bar\phi^{\text N}), P^{K}_i(\bar\phi^{\text N})\}$ represent $\{\psi^{K}_i, P^{K}_i\}$ through a Poisson isomorphism \cite{reduced1}, and thus coordinatize a reduce phase space of the physical stetes $\mathbb D^{cl}$. Subsequently, one may choose a one parameter family of the background field values $\{\bar{\phi}^{\text N}_j(\tau)\}$ with a monotonic time function $T(\bar{\phi}_j^{\text N}(\tau))$, which defines a specific foliation of the spacetimes in $\mathbb D^{cl}$. This way, we obtain an unconstrained Hamiltonian theory in the reduced phase space, with the evolutions happening in the physical time $T(\bar{\phi}^{\text N}(\tau))$under the physical background $\{\bar\phi^{\text N}_j (\tau)\}$. Finally, the Hamiltonian dynamics may be quantized into a Schr\"odinger theory with the wave functions of the form $\Psi_{(\tau)}[\psi^{K}_i]$.

In this approach, the obtained Schr\"odinger theory is built from the classical Cauchy surfaces labeled by the background fields $\bar\phi_j^{\text N}(\tau)$, which are clearly specialized in the theory to remain classical. Also, since a reduced phase space covers only the region $\mathbb D^{cl}$, the Schr\"odinger theory could not recover the full solutions of camonical general relativity in the classical limits. These issues arise from reconciling the relative and local notion of time in the classical theory, with the absolute and global time required by the quantum theory. People considering the general covariance as fundamental \cite{reduced1} tend to regard the approach as an effective approach, applicable only when the quantum fluctuations of the clock fields are negligible. On the other hand, people regarding the principle of quantum mechanics as fundamental \cite{reduced2}\cite{reduced3} tend to regard the approach as fundamental, selecting a truly physical sector of states accompanied with a privileged absolute time, which then leads to a familiar well-defined quantum field theory. For the context of this paper, we observe that the physical time in one of such these theories, absolute or not, is very different from the Newtonian time. It is given by the non-dynamical background fields which are experimentally observable just like the dynamical fields. Thus one may always ask whether the theory may still encode the possibly observable quantum nature of the background fields, in some alternative notion of time.

The second type of theories \cite{Dirac1}\cite{Dirac2} seeks to quantize the full canonical general relativity following the Dirac quantization program, such that the general covariance remains in the quantum level. Currently, the most concrete way to implement the program is given by a procedure known as the refined algebraic quantization \cite{ref1}\cite{ref2}\cite{ref3}.

In these theories, the canonical quantization is first applied to the unconstrained phase space, such that the mentioned fields are equally represented by the operators $\{\hat{\phi}^{\text N}_i, \hat{P}^{\text N}_i, \hat{\psi}^{K}_i, \hat{P}^{K}_i\}$ defined in a kinematic Hilbert space $\mathbb K$. The constraints should be also properly quantized as self-adjoint operators $\{\hat{C}^{\text N}_i\}$ in $\mathbb K$. The physical Hilbert space $\mathbb H$ is then given by the (generalized) kernel of the constraint operators. Note that $\mathbb H$ demands a definition of Hermitian inner product without an absolute notion of time. Remarkably, this inner product can be naturally defined through the refined algebraic quantization \cite{ref1}\cite{ref2}\cite{ref3}, such that the self-adjoint operators in $\mathbb K$ commuting with $\{\hat{C}^{\text N}_i\}$ become the Dirac observables in $\mathbb H$. This way, each state $\Psi \in \mathbb H$ is a quantum representation of a spacetime.

This type of theories faces many challenges from its unconventional timeless nature. For the dynamics, one has to identify the Dirac observables carrying the information of physical evolutions. A widely studied method is to identify the specific Dirac observables $\{\hat{\psi}^{K}_{i}(\bar{\phi}_j^{\text N}), \hat{P}^{K}_{i}(\bar{\phi}_j^{\text N})\}$ representing the classical observables $\{\psi^{K}_i(\bar{\phi}_j^{\text N}), P^{K}_i(\bar{\phi}_j^{\text N})\}$ mentioned above. These observables are called the quantum relational observables \cite{quantrelat1}\cite{quantrelat2}\cite{quantrelat3}, and they differ from their counter parts in the reduced phase space theories in two crucial aspects. First, one set of quantum relational observables $\{\hat{\psi}^{K}_{i}(\bar{\phi}_j^{\text N}), \hat{P}^{K}_{i}(\bar{\phi}_j^{\text N})\}$ is at the same footing of another set $\{\hat{\psi}^{'K}_{i}(\bar{\phi}_j^{'\text N}), \hat{P}^{'K}_{i}(\bar{\phi}_j^{'\text N})\}$; they are just two different sets of observables using different clocks. Therefore, the notion of time here is truly relative and compatible with the general covariance.  Second, the observables $\{\hat{\psi}^{K}_{i}(\bar{\phi}_j^{\text N}), \hat{P}^{K}_{i}(\bar{\phi}_j^{\text N})\}$ are composite operators utilizing the clock fields as the quantum field operators $\hat{\phi}^{\text N}_j$ rather than a classical background. This means that the algebra among $\{\hat{\psi}^{K}_{i}(\bar{\phi}_j^{\text N}), \hat{P}^{K}_{i}(\bar{\phi}_j^{\text N})\}$ would contain the complicated corrections from the quantum fluctuations of the clock fields. The corrections can give the quantum effects of the instruments in a realistic measurement \cite{quantrelat1}, but they also obscure the notion of a complete set of observables for the quantum theory. In contrast to the case of the reduced phase space approach, the Cauchy surfaces identified by the quantum relational observables fluctuate with the quantized clock fields, and they are unsuitable for defining a Schr\"odinger theory.
 
The goal of this paper is to propose an exact notion of Cauchy surfaces in the Dirac theories, which is analogous to that in the reduced phase space theory and suitable for defining a Schr\"odinger theory. By demands of the Dirac theories, these Cauchy surfaces must be defined at the quantum level in a relative way. In a general prescription, we will provide a natural definition for the quantum Cauchy surfaces in the context of the refined algebraic quantization. Then, we will develop a local construction of quantum Cauchy surfaces with a specified physical background for a physical subspace $\mathbb D \subset \mathbb H$. Then, we will extrapolate the conditions for a set of quantum Cauchy surfaces to form a foliation for a Schr\"odinger theory, under a specified background appearing to be without quantum fluctuations. Also, we will relate our formulation to the path integral transition amplitudes from the covariant theories of quantum gravity. Lastly, as an initial trial, we will apply the idea to a simple model with a single scalar constraint, and demonstrate that the Schr\"odinger theories of the different foliations could emerge from the same underlying Dirac theory, as perceived from the observer frames with the different backgrounds.

\section{General Prescription}
 
Refined algebraic quantization \cite{ref1}\cite{ref2}\cite{ref3} is a concrete procedure to realize the Dirac quantization of a system with first-class constraints. The application of the procedure to canonical general relativity has been a subject under intense studies \cite{lqg1}\cite{lqg2}\cite{lqg3}. In quantum cosmology, refined algebraic quantization has been completed in many symmetrically reduced models \cite{reflqc1}\cite{reflqc2}, yielding rich implications of the quantum gravitational effects. As for the full theory, the most promising on-going program lies in loop quantum gravity \cite{lqg1}\cite{lqg2}\cite{lqg3}. In the following, we summarize the steps of the procedure. 
Taking both cosmological models and the full theory into consideration, we set the ranges of the indices in $\{C^{\text N}_i\}$ to vary with the assumed symmetries. For instance, a homogeneous and isotropic model has only the scalar constraint $\text{N}=0$ with a global spatial mode $i=0$. The first step is the canonical quantization of the unconstrained phase space, leading to the kinematic Hilbert space $\mathbb K$, where the constraints are represented by a set of self-adjoint operators $\{\hat{C}^{\text N}_i\}$. Then, one must impose the quantum constraints to obtain the physical Hilbert space. This crucial step in the procedure is through finding a constraint kernel projection map $\hat{\mathbb P}:\mathbb K\to \mathbb K^*$, called a rigging map \cite{ref1}\cite{ref2}\cite{ref3}. The map has to properly implement the form
\begin{eqnarray}
\label{rigging}
\hat{\mathbb{P}}\sim \prod_{{\text N},i} \delta(\hat{C}^{\text N}_i)\,,\nonumber
\end{eqnarray}
such that its image satisfies $\hat{C}^{\text N}_i|_{Image[\hat{\mathbb P}]}=0$ in a proper sense. When this map is found, it naturally endows $Image[\hat{\mathbb P}]$ with a Hermitian inner product. Such an inner product between two physical states $|\Psi_1)=\hat{\mathbb{P}}\ket{\psi_1}$ and $|\Psi_2)=\hat{\mathbb{P}}\ket{\psi_2}$, would be given by \cite{ref1}\cite{ref2}\cite{ref3}
\begin{eqnarray}
\label{inner product}
(\Psi_1|\Psi_2)= \bra{\psi_1}\hat{\mathbb{P}}\ket{\psi_2}.
\end{eqnarray}
The physical Hilbert space $\mathbb H \equiv Image[\hat{\mathbb P}]$ is thus defined through the rigging map, and a physical state $|\Psi) \in \mathbb{H}$ is a quantum representation of a spacetime. Furthermore, with this inner product, a self-adjoint operator in $\mathbb K$ commuting with $\{ \hat{C}^{\text N}_i\}$ automatically becomes a Dirac observable \cite{ref1}\cite{ref2}\cite{ref3} in $\mathbb H$. 

For the simplest cases of the FRW quantum cosmological models with a massless scalar matter field $T$ \cite{reflqc2}\cite{reflqc3}\cite{reflqc4}, the kinematic phase space coordinates consist of the scalar-field conjugate pair $(\hat{T}, \hat{P}_T)$ and the gravitational conjugate pair $(\hat{p}\,,\,\hat{c})$ of the scale factor and the extrinsic curvature. The only constraint operator is the global scalar constraint operator $\hat{C}^{0}_{i=0}\equiv \hat C$, generating a one-parameter group of unitary transformations. In this case, $\hat{\mathbb P}$ is shown \cite{reflqc4} to be given by the group averaging expression
\begin{equation}
\begin{split}
\label{group ave}
\hat{\mathbb{P}}\equiv \int\, d\lambda \,\hat{U}(\lambda) \,;\,\,\hat{U}(\lambda) \equiv e^{-i \lambda \,\hat{C}}.
\end{split}
\end{equation}
From the resulted physical Hilbert space, various important implications about the cosmic evolution have been derived using the quantum relational observables of the form \cite{reflqc4}  
\begin{equation}
\begin{split}
\label{relational observ}
\hat{O}(\bar T)\equiv \int \,d\lambda \,\,\hat{U}(\lambda) \,\,sym\left\{ \hat{O}\,\,f(\hat{P}_T)\,\delta(\hat{T}-\bar T)\right\}\,\,\hat{U}^{-1}(\lambda)\,\,;\,\,\hat{O}\equiv O(\hat{p}\,,\,\hat{c}),
\end{split}
\end{equation}
where $sym$ denotes a proper self-adjoint symmetrization, and the factor $f(\hat{P}_T)$ serves as the absolute value of the Jacobian between $d T$ and $d\lambda$. These Dirac observables represent the value of the gravitational variables at the clock time $T=\bar T$ in the universe.

Constructing a rigging map for the full theory requires the detailed knowledge of the group of transformations generated by the scalar and momentum constraints, which have a complicated algebra with structure functions. Despite of the challenge, remarkable progresses have been achieved especially in the framework of loop quantum gravity \cite{lqg1}\cite{intro1}\cite{intro}.

As a remarkable  triumph, a rigging map solving the momentum constraints has been rigorously implemented in loop quantum gravity, leading to a spatial-diffeomorphism invariant kinematic Hilbert space \cite{area}\cite{volume}\cite{volume1} that concretely realizes the spatial quantum geometry. Recently, there are also significant advances in the pursuit of anomaly-free quantization of both the scalar and momentum constraints \cite{anomfree1}\cite{anomfree2}. Therefore, it is hopeful that a full rigging map for both the quantum scalar and momentum constraints may be defined, once the quantum constraints' algebra could be controlled and simplified. The method of the master constraint \cite{master1}\cite{master2} provides another possible direction. The classical master constraint is given by the weighted sum over the square of each of the original constraints, which combines the infinitely many original constraints into a single one. The master constraint operator in loop quantum gravity \cite{master1}\cite{master2} has been successfully constructed. Also, it has been shown \cite{master3} that the master constraint may have the proper spectrum for a rigging map. In this method with only one constraint operator, it is even hopeful that the rigging map may be constructed with the group averaging method \eqref{group ave}. 

In a broader context, the matrix elements of a rigging map defined in \eqref{inner product} have been studied as the transition amplitudes in many path integral formalisms of quantum gravity \cite{path1}\cite{path2}\cite{foam1}\cite{foam2}. Conversely, one of such consistent path integral theory may define a rigging map for a corresponding Dirac theory \cite{foam1}\cite{foam2}. Through the progress of these path integral formalisms, it is hopeful that we may understand and calculate the rigging maps via the path integral transition amplitudes. This is especially the case in loop quantum gravity, which has a path integral formalism called the spinfoam models \cite{foam1}\cite{foam2}\cite{Dirac1}. The models are originated to calculate a rigging map for the canonical loop quantum gravity, with the path integrals defined according to the actions of the constraint operators. The transition amplitudes in the models are given by summing over the descretized history of the quantum geometry \cite{foam1}\cite{foam2}\cite{Dirac1}, which may be computed perturbatively. With the remarkable progresses in the spinfoam models, it is hopeful that the models would effectively define the desired rigging map for canonical loop quantum gravity.

Having mentioned the above, we should assume that the refined algebraic quantization can be applied to quantize a theory of our concern. The resulted quantum theory is thus equipped with the triplet $(\mathbb{H}, \mathbb{K}, \hat{\mathbb{P}})$ specified above.

\subsection{Definition}

We start by making an observation in the FRW cosmology, where $\hat{\mathbb{P}}$ can be constructed by the group averaging method as in \eqref{group ave}. Recall in these cases we have the standard quantum relational observables \eqref{relational observ}, whose actions on a quantum state of spacetime $\Psi\in \mathbb H$ would be 
\begin{eqnarray}
\label{relational observ 2}
\hat{O}(\bar T) \ket{\Psi} 
&\equiv& \int \,d\lambda \,\,\hat{U}(\lambda) \,\,sym\left\{\hat{O}\,\,f(\hat{P}_T) \,\delta(\hat{T}-\bar T)\right\}\,\,\hat{U}^{-1}(\lambda)\,\,\ket{\Psi}\nonumber\\
&=&\hat{\mathbb{P}}\,\cdot\,sym\left\{\hat{O}\,\,f(\hat{P}_T) \,\delta(\hat{T}-\bar T)\right\}\ket{\Psi}
\end{eqnarray}
where the invariance $\hat{U}^{-1}(\lambda)\ket{\Psi}= \ket{\Psi}$ holds for the physical state. A classical Cauchy surface in a spacetime represents a causal instance, which contains complete but non-redundant information about the spacetime. Thus If the clock ${T}$ is to specify a unique Cauchy surface with $T=\bar T$ for every physical solution, the condition ${1}(\bar{T})= 1$ should be satisfied on the constraint surface. The analogous condition at the quantum level is
\begin{eqnarray}
\label{relational observ 3}
\hat{\bold 1}(\bar T) \ket{\Psi} 
= \hat{\mathbb{P}}\,  \cdot\,\,sym\left\{f(\hat{P}_T) \,\delta(\hat{T}-\bar T)\right\}\,\,\ket{\Psi}
= \ket{\Psi}
\end{eqnarray}
for any $\Psi\in \mathbb H$. That is, the operator definining the Cauchy surface is a right inverse operator to $\hat{\mathbb{P}}$.  

In light of this observation, we now take \eqref{relational observ 3} to be the fundamental definition of a quantum Cauchy surface. That is, for any given rigging map $\hat{\mathbb{P}}$, we define a quantum Cauchy surface $\hat{\Pi}_{t_0}$, labeled by $t_0$, to be a linear map satisfying
\begin{eqnarray}
\label{surface def}
\hat{\Pi}_{t_0}: \mathbb{H} \to \mathbb{K}\,\,\,;\,\,\,\hat{\mathbb{P}} \cdot\hat{\Pi}_{t_0}=\hat I.
\end{eqnarray}
Essentially, $\hat{\Pi}_{t_0}$ is a specific isomorphism between $\mathbb H$ and $Image[\hat{\Pi}_{t_0}]\equiv \mathbb{V}^{t_0}\subset \mathbb{K}$, so that each state $\Psi \in \mathbb{H}$ is uniquely represented by an element $\hat{\Pi}_{t_0}\Psi \equiv \psi_{t_0} \in \mathbb{V}^{t_0}$.

We denote the set of all linear operators in $\mathbb K$ as $\{\hat{O}_n:\mathbb K\to\mathbb K\}$, and the set of all Dirac observables as $\{\hat{O}_{ N}= \hat{O}^{\dagger}_{ N}:\mathbb{H}\to\mathbb{H} \}$. For a given $\hat{\Pi}_{t_0}$, denote the set of all $\mathbb{V}^{t_0}$- preserving operators as $\{\hat{O}^{{{\mathbb{V}^{{t_0}}}}}_{ m}\}\subset \{\hat{O}_n\}$. Through the isomorphism, the Dirac observables $\{\hat{O}_{ N}\}$ can be represented by a corresponding set of operators $\{\hat{O}^{{{\mathbb{V}^{{t_0}}}}}_{{\mathcal N}}\}\subset \{\hat{O}^{{{\mathbb{V}^{{t_0}}}}}_{ m}\}$ satisfying
\begin{eqnarray}
\label{Dirac observ 1}
\hat{O}_{{\bar N}}= \hat{\mathbb{P}}\,\,\hat{O}^{{{\mathbb{V}^{{t_0}}}}}_{{\bar{\mathcal N}}}\,\,\hat{\Pi}_{t_0}\equiv \hat{O}^{{{\mathbb{V}^{{t_0}}}}}_{{\bar{\mathcal N}}}(t_0).
\end{eqnarray}
Clearly from the definition, the restriction $\hat{O}^{{{\mathbb{V}^{{t_0}}}}}_{{\bar{\mathcal N}}}|_{\mathbb{V}^{{t_0}}}$ is determined by $\hat{O}_{{\bar N}}$, while its extension outside of the subspace $\mathbb{V}^{{t_0}}$ is arbitrary. Also, we have introduced the notation $\hat{O}^{{{\mathbb{V}^{{t_0}}}}}_{{\bar{\mathcal N}}}(t_0)$ with the interpretation that $\hat{O}_{{\bar N}}$ represents the value of ${O}^{{{\mathbb{V}^{{t_0}}}}}_{{\bar{\mathcal N}}}$ at the Cauchy surface $t_0$. This interpretation is supported by the algebraic isomorphism 
\begin{eqnarray}
\label{isomorphism1}
\hat{O}^{{{\mathbb{V}^{{t_0}}}}}_{{{\mathcal N}}}(t_0) \,\hat{O}^{{{\mathbb{V}^{{t_0}}}}}_{{{\mathcal N'}}}(t_0)= \left(\hat{O}^{{{\mathbb{V}^{{t_0}}}}}_{{{\mathcal N}}}\,\hat{O}^{{{\mathbb{V}^{{t_0}}}}}_{{{\mathcal N'}}}\right)(t_0) \,\,\,\text{and}\,\,\,\, \left[\hat{O}^{{{\mathbb{V}^{{t_0}}}}}_{{{\mathcal N}}}(t_0), \hat{O}^{{{\mathbb{V}^{{t_0}}}}}_{{{\mathcal N'}}}(t_0)\right]= \left[\hat{O}^{{{\mathbb{V}^{{t_0}}}}}_{{{\mathcal N}}}, \hat{O}^{{{\mathbb{V}^{{t_0}}}}}_{{{\mathcal N'}}}\right](t_0).
\end{eqnarray}
To understand the classical limits of the observables $\{\hat{O}^{{{\mathbb{V}^{{t_0}}}}}_{{{\mathcal N}}}(t_0)\}$, it is necessary to relate them to the self-adjoint subset $\{\hat{O}^{{{\mathbb{V}^{{t_0}}}}}_{ N}=\hat{O}^{ {{\mathbb{V}^{{t_0}}}}\dagger}_{ N}\}\subset\{\hat{O}^{{{\mathbb{V}^{{t_0}}}}}_{{m}}\}$. Note that the operator $\hat{O}^{{{\mathbb{V}^{{t_0}}}}}_{{\bar{\mathcal N}}}$ in \eqref{Dirac observ 1} generally cannot be chosen as self-adjoint in $\mathbb K$, since $\hat{\Pi}_{t_0}$ is generally not an isometry between the two Hilbert spaces $\mathbb{V}^{t_0}$ and $\mathbb{H}$. However, $\hat{\Pi}_{t_0}$ is always related to an isometry through a similar transformation in $\mathbb{V}^{t_0}$. Setting $\{|E_k)\}$ to be an orthonormal basis of $\mathbb{H}$, we introduce a similar map $\hat{\Lambda}_{t_0}\in \{\hat{O}^{{{\mathbb{V}^{{t_0}}}}}_{ m}\}$ such that $\{\ket{e_K}\equiv \hat{\Lambda}_{t_0}\hat{\Pi}_{t_0}|E_k) \}$ gives an orthonormal basis of $\mathbb{V}^{t_0}$. The isometry $\hat{\Lambda}_{t_0}\hat{\Pi}_{t_0}$ then leads to a new expression of \eqref{Dirac observ 1} with
\begin{eqnarray}
\label{Dirac observ 2}
\hat{O}^{{{\mathbb{V}^{{t_0}}}}}_{{\bar{\mathcal N}}} = \hat{\Lambda}_{t_0}^{\!-\!1}\hat{O}^{{{\mathbb{V}^{{t_0}}}}}_{{\bar N}}\hat{\Lambda}_{t_0}\,\,; \,\, \hat{O}^{{{\mathbb{V}^{{t_0}}}}}_{{\bar{\mathcal N}}}(t_0) =\left(\hat{\Lambda}_{t_0}^{\!-\!1}\hat{O}^{{{\mathbb{V}^{{t_0}}}}}_{{\bar N}}\hat{\Lambda}_{t_0}\right)(t_0). 
\end{eqnarray}
Here, $\hat{O}^{{{\mathbb{V}^{{t_0}}}}}_{{\bar N}}|_{\mathbb{V}^{{t_0}}}$ is self-adjoint and determined up to an unitary ambiguity in choosing the $\hat{\Lambda}_{t_0}$. The algebraic isomorphism above is then carried over as  
\begin{eqnarray}
\label{isomorphism2}
(\hat{\Lambda}_{t_0}^{\!-\!1}\hat{O}^{{{\mathbb{V}^{{t_0}}}}}_{{ N}}\hat{\Lambda}_{t_0})(t_0) (\hat{\Lambda}_{t_0}^{\!-\!1}\hat{O}^{{{\mathbb{V}^{{t_0}}}}}_{{ N}'}\hat{\Lambda}_{t_0})(t_0) = \left(\hat{\Lambda}_{t_0}^{\!-\!1}\hat{O}^{{{\mathbb{V}^{{t_0}}}}}_{{ N}}\hat{O}^{{{\mathbb{V}^{{t_0}}}}}_{{ N}'}\hat{\Lambda}_{t_0}\right)(t_0)\,,\,\,\text{and}
\nonumber\\
\,\,\,\, \left[(\hat{\Lambda}_{t_0}^{\!-\!1}\hat{O}^{{{\mathbb{V}^{{t_0}}}}}_{{ N}}\hat{\Lambda}_{t_0})(t_0) ,(\hat{\Lambda}_{t_0}^{\!-\!1}\hat{O}^{{{\mathbb{V}^{{t_0}}}}}_{{ N}'}\hat{\Lambda}_{t_0})(t_0) \right]=\left(\hat{\Lambda}_{t_0}^{\!-\!1}\left[\hat{O}^{{{\mathbb{V}^{{t_0}}}}}_{{ N}},\hat{O}^{{{\mathbb{V}^{{t_0}}}}}_{{ N}'}\right]\hat{\Lambda}_{t_0}\right)(t_0).
\end{eqnarray}
In the reverse direction, we have the following important conclusion. Given a quantum Cauchy surface $\hat{\Pi}_{t_0}$, each complete set of self-adjoint of operators in $\mathbb{V}^{t_0}$, namely the set $\{\hat{O}^{{{\mathbb{V}^{{t_0}}}}}_{{N}}\}$, determines a complete set of Dirac observables $\{(\hat{\Lambda}_{t_0}^{\!-\!1}\hat{O}^{{{\mathbb{V}^{{t_0}}}}}_{{ N}}\hat{\Lambda}_{t_0})(t_0)\}$ up to an unitary ambiguity in $\mathbb H$ associated to the choice of $\hat{\Lambda}_{t_0}$.

Let us now observe the relation between the quantum relational observables and the Dirac observables we just defined. We look into the simple case of the FRW quantum cosmology by comparing the expression \eqref{relational observ 2} with the expression \eqref{Dirac observ 1} when applied to the setting. We see that the two types of observables may be closely related if the pair $(\,f(\hat{P}_T)\,\delta(\hat{T}-\bar T)\,,\,O(\hat{p}\,,\,\hat{c})\,)$ is identifiable with the pair $(\,\hat{\Pi}_{t_0}\,,\,\hat{\Lambda}_{t_0}^{-1}\hat{O}^{{{\mathbb{V}^{{t_0}}}}}_{{\bar N}}\hat{\Lambda}_{t_0}\,)$ under certain approximations. In many interesting models where the $T$ field has a good asymptotic behavior, the condition \eqref{relational observ 3} would hold exactly \cite{reflqc4}\cite{marolf} with a proper $f(\hat{P}_T)$. In these models we may choose $\hat{\Pi}_{t_0}\equiv sym\{f(\hat{P}_T)\,\delta(\hat{T}-\bar T)\}$ as a quantum Cauchy surface. The corresponding subspace $\mathbb{V}^{{t_0}}$ would be an eigenspace of $\hat T$ with $\hat T|_{\mathbb{V}^{{t_0}}}=\bar T$, and it would also have a complete set of self-conjugate operators $\{{O}^{{{\mathbb{V}^{{t_0}}}}}_{ N}(\hat c,  \hat p)\}$. With each given $\hat{\Lambda}_{t_0}$, the set gives a complete set of Dirac observables $\{(\hat{\Lambda}_{t_0}^{\!-\!1}{O}^{{{\mathbb{V}^{{t_0}}}}}_{{ N}}(\hat c,  \hat p)\hat{\Lambda}_{t_0})(t_0)\}$. When the given $\hat{\Lambda}_{t_0}$ is such that $(\hat{\Lambda}_{t_0}^{\!-\!1}{O}^{{{\mathbb{V}^{{t_0}}}}}_{{ N}}(\hat c,  \hat p)\hat{\Lambda}_{t_0})={O}^{{{\mathbb{V}^{{t_0}}}}}_{{N}}(\hat c,  \hat p)+\mathcal{O}(\hbar)$, we have  
\begin{eqnarray}
\label{Dirac observ 2}
(\hat{\Lambda}_{t_0}^{\!-\!1}\hat{O}^{{{\mathbb{V}^{{t_0}}}}}_{{ N}}\hat{\Lambda}_{t_0})(t_0)=\hat{O}^{{{\mathbb{V}^{{t_0}}}}}_{{ N}}(\bar T)+\mathcal{O}(\hbar) \,\,;\,\,\, (\hat{\Lambda}_{t_0}^{-1}\,\, \hat{T}\,\,\hat{\Lambda}_{t_0})(t_0)={\bar T}. 
\end{eqnarray}
Therefore, in these models the two sets of observables are two distinct quantum representations of the same classical observables.

Here, the difference term of $\mathcal{O}(\hbar)$ represents a crucial distinction between the two types of observables. The self adjointness of the quantum relational observables is established through the self-adjoint extensions, which symmetrize the operator orderings. On the other hand, our observables are isomorphically induced from the set of kinematic self-adjoint operators in $\mathbb{V}^{{t_0}}$. Therefore, while the relational quantum observables' algebra contains the quantum corrections from the quantum clocks, the quantum Cauchy surfaces lead to the Dirac observables faithfully representing their kinematic counterparts.

\subsection{Local Construction of a Quantum Cauchy Surface}

The purpose of introducing the quantum Cauchy surfaces is to derive a Schr\"odinger theory from a Dirac theory $(\mathbb{H}, \mathbb{K}, \hat{\mathbb{P}})$. To do that, we first identify the instantaneous wave functions at a quantum Cauchy surface with a given physical background. 

Let us continue using the previous notations with $\mathbb K \equiv Span\{\ket{\phi^{\text N}_i}\otimes\ket{\psi^{K}_i}\}$, in which the field operators $\{\hat{\phi}^{\text N}_i, \hat{P}^{\text N}_i, \hat{\psi}^{K}_i, \hat{P}^{K}_i\}$ are defined. Although all the fields are quantized in the Dirac theory, the physical quantum degrees of freedom are ``fewer" than those of $\mathbb K$ due to the quantum constraints $\{\hat C^{\text N}_i\}$. By our construction, these physical degrees of freedom are exactly the degrees of freedom in $\mathbb{V}^{{\tau_0}}$. Our goal is to further show that the degrees of freedom in $\mathbb{V}^{{\tau_0}}$ can provide the physical spectra to certain Schr\"odinger wave functions, at the moment of time given by the quantum Cauchy surface $\hat{\Pi}_{\tau_0}$. Such a moment could be labeled by a certain set of background fields, whose quantum fluctuations are absent in $\mathbb{V}^{{\tau_0}}$ due to the constraints. They would be as many as the number of the constraints, and taking a set of specific eigenvalues in $\mathbb{V}^{{\tau_0}}$. For a set of chosen background fields, we expect such a description to be valid only locally in a sub-Hilbert space $\mathbb D\subset\mathbb H$. Accordingly, we would also localize our construction in the following way.

We now set the instantaneous background to be $\phi^{\text N}_i=\bar{\phi}^{\text N}_i(\tau_0)$. This restricts $\mathbb K$ to its maximal subspace consistent with the background, which is just the corresponding eigenspacce given by ${\mathbb S}_{\star}^{\tau_0} \equiv Span\{\ket{\bar{\phi}^{\text N}_i(\tau_0)}\otimes\ket{\psi^{K}_i}\}$. This also defines the domain of the Heisenberg states for our Schr\"odinger theory, which would be $\mathbb D_{\tau_0}\equiv \hat{\mathbb P}\,{\mathbb S}_{\star}^{\tau_0} \subset\mathbb H$. In general, the spectra of the wave functions cannot be found in ${\mathbb S}_{\star}^{\tau_0}$, as it may have a nontrivial kernel of $\hat{\mathbb P}$ and thus fail to be isomorphic to $\mathbb D_{\tau_0}$. Instead, we should look into a quantum Cauchy surface $\hat{\Pi}_{\tau_0}$ with $\mathbb S^{\tau_0}\equiv\hat{\Pi}_{\tau_0}\mathbb{D}_{\tau_0} \subset {\mathbb S}_{\star}^{\tau_0}$. Being consistent with the background and isomorphic to ${\mathbb D}_{\tau_0}$, the subspace $\mathbb S^{\tau_0}$ would satisfy
\begin{eqnarray}
\label{compliment}
\mathbb S^{\tau_0}\oplus Ker[\hat{\mathbb{P}}|_{{\mathbb S}_{\star}^{\tau_0}}]= {\mathbb S}_{\star}^{\tau_0}.
\end{eqnarray}
Reversely, each $\mathbb S^{\tau_0}$ satisfying the above defines a quantum Cauchy surface in the domain $\mathbb D_{\tau_0}$ which is consistent with the background $\phi^{\text N}_i=\bar{\phi}^{\text N}_i(\tau_0)$. 

Therefore, the first important step is identifying the kernel $Ker[\hat{\mathbb{P}}|_{{\mathbb S}_{\star}^{\tau_0}}]$. The kernel is given by the set of elements corresponding to the zero vector in $\mathbb H$ with zero norm under the rigging map. Since the rigging map also defines the inner product through \eqref{inner product}, the basis of the kernel is specified by the linearly independent maximal set $ \{\ket{{e^{\,{}_{{\mathbb S}_{\star}^{\tau_0}}}}_{\!\!\lambda}}\} \subset {\mathbb S}_{\star}^{\tau_0}$ (with the members labeled by $\lambda$) satisfying 
\begin{eqnarray}
\label{zero kernel}
\bra{{e^{\,{}_{{\mathbb S}_{\star}^{\tau_0}}}}_{\!\!\lambda}}\hat{\mathbb{P}} \ket{{e^{\,{}_{{\mathbb S}_{\star}^{\tau_0}}}}_{\!\!\lambda}}= 0\,.
\end{eqnarray}
The space $Span\{\ket{{e^{\,{}_{{\mathbb S}_{\star}^{\tau_0}}}}_{\!\!\lambda}}\}=Ker[\hat{\mathbb{P}}|_{{\mathbb S}_{\star}^{\tau_0}}]$ then specify the valid choices for $\mathbb S^{\tau_0}$ satisfying \eqref{compliment}. Clearly whenever the kernel is non-trivial, the space $\mathbb S^{\tau_0}$ is non-unique.

Suppose the kernel is determined and a specific $\mathbb S^{\tau_0}$ is chosen correspondingly. This gives us a quantum Cauchy surface $\hat{\Pi}_{\tau_0}$ specified in ${\mathbb D}_{\tau_0}$. To define the wave functions we are looking for, we use \eqref{Dirac observ 1} to construct the Dirac observables preserving the Heisenberg domain $\mathbb D_{\tau_0}$. By definition, they are the representations of the $\mathbb S^{\tau_0}$-preserving set $\{\hat{O}^{{{\mathbb S^{\tau_0}}}}_{{\bold N}}\}\subset \{\hat{O}^{{{\mathbb V^{\tau_0}}}}_{{ N}}\}$, whose restriction $\{\hat{O}^{{{\mathbb S^{\tau_0}}}}_{{\bold N}}|_{\mathbb S^{\tau_0}}\equiv {\bold O}^{{{\mathbb S^{\tau_0}}}}_{{ \bold N}}(\hat{\phi}^{\text N}_i,\hat{\psi}^{K}_i, \hat{P}^{K}_i):\mathbb S^{\tau_0} \to \mathbb S^{\tau_0}\}$ is the set of all the self-adjoint operators in $\mathbb S^{\tau_0}$. 

The observables also require a $\mathbb S^{\tau_0}$-preserving $\hat{\Lambda}_{\tau_0}$, which will be denoted as $\hat{\Lambda}_{\mathbb S^{\tau_0}}$, whose restriction $\hat{\Lambda}_{\mathbb S^{\tau_0}}|_{\mathbb S^{\tau_0}}\equiv \hat{\bold\Lambda}_{\mathbb S^{\tau_0}}:\mathbb S^{\tau_0} \to \mathbb S^{\tau_0}$ is a similar transformation in $\mathbb S^{\tau_0}$. Set $\{\ket{e^{{}_{{\mathbb S}^{\tau_0}}}_{K}}\}$ to be an orthonormal basis of the Hilbert space $\mathbb S^{\tau_0}$. In this paper, we assume integer values for the index $K$, for the convenience of our description. By definition, the basis $\{ \,\hat{\mathbb{P}}\,\hat{\bold\Lambda}^{\!-\!1}_{\mathbb S^{\tau_0}}\ket{e^{{}_{{\mathbb S}^{\tau_0}}}_{K}}\equiv |E^{{}_{{\mathbb S}^{\tau_0}}}_{K}) \}$ of $\mathbb D_{\tau_0}$ should be orthonormal in the Hilbert space $\mathbb H$. This condition then defines $\hat{\bold\Lambda}_{\mathbb S^{\tau_0}}$ through the equation
\begin{eqnarray}
\label{Lambda1}
 {({\bold\Lambda}^{\!-\!1}_{\mathbb S^{\tau_0}})^*}_L^J\,\,({\bold\Lambda}^{\!-\!1}_{\mathbb S^{\tau_0}})_M^K \,\,\bra{e^{{}_{{\mathbb S}^{\tau_0}}}_{J}}\hat{\mathbb{P}}\ket{e^{{}_{{\mathbb S}^{\tau_0}}}_{K}}
= (E^{{}_{{\mathbb S}^{\tau_0}}}_{L}|E^{{}_{{\mathbb S}^{\tau_0}}}_{M})=\delta_{L,M}.
\end{eqnarray}
Obviously, the $\hat{\bold\Lambda}_{\mathbb S^{\tau_0}}$ is defined up to a unitary transformation in $\mathbb S^{\tau_0}$.

So far we have argued that, given a background $\phi^{\text N}_i=\bar{\phi}^{\text N}_i(\tau_0)$, via \eqref{zero kernel} and \eqref{Lambda1} the rigging map elements in ${\mathbb S}_{\star}^{\tau_0}$ define a quantum Cauchy surface $\hat{\Pi}_{\tau_0}$ and the corresponding Dirac observables in $\mathbb D_{\tau_0}$, up to the two ambiguities. Before addressing the ambiguities, let us write down the instantaneous wave functions.

Through \eqref{Dirac observ 1}, the set of $\mathbb D_{\tau_0}$ preserving Dirac observables is the set $\{(\hat{\Lambda}_{\mathbb S^{\tau_0}}^{\!-\!1}\hat{O}^{{\mathbb{S}^{\tau_0}}}_{{\bold N}}\hat{\Lambda}_{\mathbb S^{\tau_0}})(\tau_0)\}$, whose action on a state $\Psi_{\mathbb D_{\tau_0}}\in \mathbb D_{\tau_0}$ takes the form
\begin{eqnarray}
\label{0}
 (\hat{\Lambda}_{\mathbb S^{\tau_0}}^{\!-\!1}\hat{O}^{{\mathbb{S}^{\tau_0}}}_{{\bold N}}\hat{\Lambda}_{\mathbb S^{\tau_0}})(\tau_0)\,\,\cdot\big|\Psi_{\mathbb D_{\tau_0}}\big)= \hat{\mathbb P}\, \,\hat{\bold\Lambda}_{\mathbb S^{\tau_0}}^{\!-\!1}\,\hat{\bold O}^{{\mathbb{S}^{\tau_0}}}_{{\bold N}}\,\hat{\bold\Lambda}_{\mathbb S^{\tau_0}}\,\,\hat{\Pi}_{\tau_0}\,\,\cdot\big|\Psi_{\mathbb D_{\tau_0}}\big).
\end{eqnarray}
Thus, each conjugate pair of complete sets in $\mathbb{S}^{\tau_0}$, consisting of $\{\hat{\bold{X}}^{{\mathbb{S}^{\tau_0}}}_{{\bold N}}\}\subset \{\hat{\bold O}^{{\mathbb{S}^{\tau_0}}}_{{\bold N}}\}$ and $\{\hat{\bold{P}}^{{\mathbb{S}^{\tau_0}}}_{{\bold N}}\}\subset \{\hat{\bold O}^{{\mathbb{S}^{\tau_0}}}_{{\bold N}}\}$, gives a conjugate pair of complete Dirac Observables in $\mathbb D_{\tau_0}$, consisting of $\{(\hat{\Lambda}_{\mathbb S^{\tau_0}}^{\!-\!1}\hat{X}^{{\mathbb{S}^{\tau_0}}}_{{\bold N}}\hat{\Lambda}_{\mathbb S^{\tau_0}})(\tau_0)\}$ and $\{(\hat{\Lambda}_{\mathbb S^{\tau_0}}^{\!-\!1}\hat{{P}}^{{\mathbb{S}^{\tau_0}}}_{{\bold N}}\hat{\Lambda}_{\mathbb S^{\tau_0}})(\tau_0)\}$. Also, we would have $\hat{\bold\Lambda}_{\mathbb S^{\tau_0}}={\bold\Lambda}_{\mathbb S^{\tau_0}}(\hat{\bold{X}}^{{\mathbb{S}^{\tau_0}}}_{{\bold N}},\hat{\bold{P}}^{{\mathbb{S}^{\tau_0}}}_{{\bold N}})$.

An orthonormal eigenbasis for $\mathbb D_{\tau_0}$ of $\{(\hat{\Lambda}_{\mathbb S^{\tau_0}}^{\!-\!1}\hat{X}^{{\mathbb{S}^{\tau_0}}}_{{\bold N}}\hat{\Lambda}_{\mathbb S^{\tau_0}})(\tau_0)\}$, denoted as $\{|{{\bold X}^{{\mathbb{S}^{\tau_0}}}_{{\bold N}}}{ ; \, \hat{\Lambda}_{\mathbb S^{\tau_0}}})\}$, can be read off directly as
\begin{eqnarray}
\label{eigenstate 1}
\big|{{\bold X}^{{\mathbb{S}^{\tau_0}}}_{{\bold N}}}{ ; \, \hat{\Lambda}_{\mathbb S^{\tau_0}}}\big)\equiv\, \hat{\mathbb P} \, \hat{\bold\Lambda}_{\mathbb S^{\tau_0}}^{\!-\!1}\,\ket{\bar{\phi}^{\text N}_i(\tau_0)\,,\,{\bold X}^{{\mathbb{S}^{\tau_0}}}_{{\bold N}}},
\end{eqnarray}
where $\{\ket{\bar{\phi}^{\text N}_i(\tau_0)\,,\,{\bold X}^{{\mathbb{S}^{\tau_0}}}_{{\bold N}}}\}$ is an orthonormal eigenbasis for $\mathbb{S}^{{\tau_0}}$ of $\{\hat{\bold{X}}^{{\mathbb{S}^{\tau_0}}}_{{\bold N}}\}$. 
It can be easily checked that the basis indeed satisfies
\begin{eqnarray}
\label{eigenstate 2}
(\hat{\Lambda}_{\mathbb S^{\tau_0}}^{\!-\!1}\hat{X}^{{\mathbb{S}^{\tau_0}}}_{{\bold N}}\hat{\Lambda}_{\mathbb S^{\tau_0}})(\tau_0)\,\cdot\,\big|{\bar{\bold X}^{{\mathbb{S}^{\tau_0}}}_{{\bold N}}}{ ; \, \hat{\Lambda}_{\mathbb S^{\tau_0}}}\big) = \bar{\bold X}^{{\mathbb{S}^{\tau_0}}}_{{\bold N}} \,\big|{\bar{\bold X}^{{\mathbb{S}^{\tau_0}}}_{{\bold N}}}{ ; \, \hat{\Lambda}_{\mathbb S^{\tau_0}}}\big)\,;\nonumber\\
(\hat{\Lambda}_{\mathbb S^{\tau_0}}^{\!-\!1}\hat{\phi}^{\text N}_i\hat{\Lambda}_{\mathbb S^{\tau_0}})(\tau_0)\,\cdot\,\big|{\bar{\bold X}^{{\mathbb{S}^{\tau_0}}}_{{\bold N}}}{ ; \, \hat{\Lambda}_{\mathbb S^{\tau_0}}}\big) = \bar{\phi}^{\text N}_i(\tau_0) \,\big|{\bar{\bold X}^{{\mathbb{S}^{\tau_0}}}_{{\bold N}}}{ ; \, \hat{\Lambda}_{\mathbb S^{\tau_0}}}\big).
\end{eqnarray}
Finally, each normalized state $\Psi_{\mathbb D_{\tau_0}}\in \mathbb D_{\tau_0}$ corresponds to a normalized wave function
\begin{eqnarray}
\label{wave function 1}
\Psi_{\mathbb D_{\tau_0}\,(\tau_0)}\big[\,{{\bold X}^{{\mathbb{S}^{\tau_0}}}_{{\bold N}}}{ ; \, \hat{\Lambda}_{\mathbb S^{\tau_0}}}\,\big] \equiv \big({{\bold X}^{{\mathbb{S}^{\tau_0}}}_{{\bold N}}}{ ; \, \hat{\Lambda}_{\mathbb S^{\tau_0}}}\big| \Psi_{\mathbb D_{\tau_0}} \big).
\end{eqnarray}
This gives the instantaneous ${\bold X}^{{\mathbb{S}^{\tau_0}}}_{{\bold N}}$-representation of $\Psi_{\mathbb D_{\tau_0}}$, under the background $\bar{\phi}^{\text N}_i(\tau_0)$. The ${\bold P}^{{\mathbb{S}^{\tau_0}}}_{{\bold N}}$-representation would then be the conjugate representation.

Finally, we address the ambiguities in the choices of $\mathbb{S}^{{\tau_0}}$ and $\hat{\Lambda}_{\mathbb S^{\tau_0}}$ under the given background $\bar{\phi}^{\text N}_i(\tau_0)$. Let us consider the possibility of another valid pair $(\mathbb{S}^{'{\tau_0}},\hat{\Lambda}_{\mathbb{S}^{{}_{'}{\tau_0}}})$. That means, $\mathbb{S}^{'{\tau_0}}$ satisfies \eqref{compliment} and we have $\{ \,\hat{\mathbb{P}}\,\!\hat{\Lambda}^{\!-\!1}_{\mathbb{S}^{{}_{'}{\tau_0}}}\ket{e^{\mathbb{S}^{{}_{'}{\tau_0}}}_{K}}= |E^{{}_{\mathbb{S}^{{}_{'}{\tau_0}}}}_{K})\equiv |E^{{}_{\mathbb{S}^{{\tau_0}}}}_{K}) \}$ for some orthonormal basis $\{\ket{e^{\mathbb{S}^{{}_{'}{\tau_0}}}_{K}}\}$ of $\mathbb{S}^{'{\tau_0}}$. Relating to the original choice, there should be an unitary operator $\hat{U}= U(\hat{\phi}^{\text N}_i,\hat{\psi}^{K}_i, \hat{P}^{K}_i): {\mathbb S}_{\star}^{\tau_0}\to {\mathbb S}_{\star}^{\tau_0}$ which transforms $\{\ket{e^{{}_{{\mathbb S}^{\tau_0}}}_{K}}\}$ into $\{\ket{e^{\mathbb{S}^{{}_{'}{\tau_0}}}_{K}}\}$. Then, using the set $(\hat{X}^{{{\mathbb{S}^{{}_{'}{\tau_0}}}}}_{{\bold N}}, \hat{{P}}^{\mathbb{S}^{{}_{'}{\tau_0}}}_{{\bold N}})\equiv(\hat{U}^{\!-\!1}\hat{X}^{{\mathbb{S}^{\tau_0}}}_{{\bold N}}\hat{U}\,, \,\hat{U}^{\!-\!1}\hat{{P}}^{{\mathbb{S}^{\tau_0}}}_{{\bold N}}\hat{U})$ one can easily show that
 \begin{eqnarray}
\label{obsv equiv}
\left(\,(\hat{\Lambda}_{\mathbb S^{\tau_0}}^{\!-\!1}\hat{X}^{{\mathbb{S}^{\tau_0}}}_{{\bold N}}\hat{\Lambda}_{\mathbb S^{\tau_0}})(\tau_0)\,,\,(\hat{\Lambda}_{\mathbb S^{\tau_0}}^{\!-\!1}\hat{{P}}^{{\mathbb{S}^{\tau_0}}}_{{\bold N}}\hat{\Lambda}_{\mathbb S^{\tau_0}})(\tau_0)\,\right)= \left(\,(\hat{\Lambda}^{\!-\!1}_{\mathbb{S}^{{}_{'}{\tau_0}}}\,\hat{X}^{\mathbb{S}^{{}_{'}\tau_0}}_{{\bold N}}\,\hat{\Lambda}_{\mathbb{S}^{{}_{'}{\tau_0}}})(\tau_0)\,,\,(\hat{\Lambda}^{\!-\!1}_{\mathbb{S}^{{}_{'}{\tau_0}}}\,\hat{{P}}^{\mathbb{S}^{{}_{'}\tau_0}}_{{\bold N}}\,\!\hat{\Lambda}_{\mathbb{S}^{{}_{'}{\tau_0}}})(\tau_0)\,\right).\nonumber\\
\end{eqnarray}
This then leads to
 \begin{eqnarray}
\label{wave function equiv}
\Psi_{\mathbb D_{\tau_0}\,(\tau_0)}\big[\,{{\bold X}^{{\mathbb{S}^{\tau_0}}}_{{\bold N}}}{ ; \, \hat{\Lambda}_{\mathbb S^{\tau_0}}}\,\big] = \Psi_{\mathbb D_{\tau_0}\,(\tau_0)}\big[\,{\bold{X}^{{{\mathbb{S}^{{}_{'}{\tau_0}}}}}_{{\bold N}}}{ ; \, \hat{\Lambda}_{\mathbb{S}^{{}_{'}{\tau_0}}}}\,\big] \bigg|_{{{\bold X}^{{\mathbb{S}^{\tau_0}}}_{{\bold N}}}=\bold{X}^{{{\mathbb{S}^{{}_{'}{\tau_0}}}}}_{{\bold N}}}.
\end{eqnarray}
It is clearly now that the different choices of $\mathbb{S}^{{\tau_0}}$ and $\hat{\Lambda}_{\mathbb S^{\tau_0}}$ are related by the changes of the variables in $\hat{\psi}^{K}_i$ and $\hat{P}^{K}_i$, in the expressions of the observables under the background $\bar{\phi}^{\text N}_i(\tau_0)$. 

\subsection{Quantum Foliation and Schr\"odinger Theories  }

To explore the dynamics of the theory, we specify a background ${\phi}^{\text N}_i=\bar{\phi}^{\text N}_i(\tau)$ over various $\tau$, which gives a scalar time field $T_j(\bar{\phi}^{\text N}_i(\tau))$ increasing monotonically with $\tau$. Following our construction in the previous section, we find a family of quantum Cauchy surfaces $\{\hat{\Pi}_{\tau}\}$, whose each member $\hat{\Pi}_{\tau}$ has the specified background $\bar{\phi}^{\text N}_i(\tau)$ and the corresponding domain ${\mathbb D}_{\tau}$. Suppose ${\mathbb D}_{\tau}= \mathbb D$, then each normalized state $\Phi_{\mathbb D} \in \mathbb D$ can be written as a normalized wave function $\Psi_{\mathbb D\,(\tau)}\big[\,{{\bold X}^{{{\mathbb{S}^{{\tau}}}}}_{{\bold N}}}{ ; \, \hat{\Lambda}_{\mathbb S^{\tau}}}\,\big]$ at any value of $\tau$. If we further have $(\hat{\bold X}^{{{\mathbb{S}^{{\tau}}}}}_{{\bold N}}, \hat{\bold P}^{{{\mathbb{S}^{{\tau}}}}}_{{\bold N}})=(\hat{\bold X}_{{\bold N}}, \hat{\bold P}_{{\bold N}})$, the the family of wave functions would share a common spectrum basis. This allows a wave function $\bold{\Psi}_{\mathbb D\,(\tau)}\big[{{\bold X}_{{\bold N}}}{ ; \, \hat{\Lambda}_{\mathbb S^{\tau}}}\big]$ describing an evolution over the spectrum of $\hat{\bold X}_{{\bold N}}$. Also, the unitary evolution of $\bold{\Psi}_{\mathbb D\,(\tau)}$ in $\tau$ would be governed by an effective self-adjoint Hamiltonian operator $\hat{\bold H}={\bold H}(\hat{\bold X}_{{\bold N}}, \hat{\bold P}_{{\bold N}} ;\tau)$. Therefore, we see that the two stability conditions lead to a Schr\"odinger theory. In the following, we address the two stability conditions separately.

 We first look into the stability of the observables $\{{\bold X}^{{{\mathbb{S}^{{\tau}}}}}_{{\bold N}},{\bold P}^{{{\mathbb{S}^{{\tau}}}}}_{{\bold N}}\}$. Let us set ${\bold X}^{{\mathbb{S}^{\tau_0}}}_{{\bold N}}\equiv {\bold X}_{{\bold N}}$, and observe that the stability is equivalent to the validity of the choice with $\mathbb{S}^{{\tau}}= Span\{\ket{\bar{\phi}^{\text N}_i(\tau)}\otimes\ket{{\bold X}^{{\mathbb{S}^{\tau_0}}}_{{\bold N}}}\}$. From the previous section, this specific form of $\mathbb{S}^{{\tau}}$ is valid if and only if the equation $\mathbb S^{\tau}\oplus Ker[\hat{\mathbb{P}}|_{{\mathbb S}_{\star}^{\tau}}]= {\mathbb S}_{\star}^{\tau}$ is satisfied. Here we know that $\mathbb{S}^{{\tau}}_\star= Span\{\ket{\bar{\phi}^{\text N}_i(\tau)}\otimes\ket{\Psi^K_i} \}$, and $Ker[\hat{\mathbb{P}}|_{{\mathbb S}_{\star}^{\tau}}]= Span\{\ket{\bar{\phi}^{\text N}_i(\tau)}\otimes\ket{\eta_{n}(\tau)}\}$ with $\{\ket{\eta_{n}(\tau)}\}$ being a basis for a $\tau$ dependent subspace of the $\Psi^K_i$-space. Therefore, the equation is equivalent to
 \begin{eqnarray}
\label{observable stability}
Span\{\ket{{\bold X}^{{\mathbb{S}^{\tau_0}}}_{{\bold N}}}\}\oplus Span\{\ket{\eta_{n}(\tau)}\} = Span\{\ket{\Psi^K_i} \}.
\end{eqnarray}
Note that the only $\tau$ dependence lies in the second term, and that the equation is satisfied at $\tau=\tau_0$ by definition.
Let $\tau$ increases from $\tau_0$, as long as the space $Span\{\ket{\eta_{n}(\tau)}\}$ varies continuously, we expect it to remain a complement to 
$Span\{\ket{{\bold X}^{{\mathbb{S}^{\tau_0}}}_{{\bold N}}}\}$ in $Span\{\ket{\Psi^K_i} \}$ in a finite range $\tau_0\leq \tau\leq \tau'_1$. Within that range, we may set ${\bold X}^{{{\mathbb{S}^{{\tau}}}}}_{{\bold N}}\equiv {\bold X}^{{\mathbb{S}^{\tau_0}}}_{{\bold N}}\equiv {\bold X}_{{\bold N}}$ and ${\bold P}^{{{\mathbb{S}^{{\tau}}}}}_{{\bold N}}\equiv {\bold P}^{{\mathbb{S}^{\tau_0}}}_{{\bold N}}\equiv {\bold P}_{{\bold N}}$, thereby obtaining a stable set of observables in the range $\tau_0\leq \tau \leq\tau'_1$.

Now we look into the domain stability $\mathbb D_\tau=\mathbb D$, which says that the quantum Cauchy surfaces must define a foliation for a definite set of quantum spacetimes given by $\mathbb D$. Such a condition has to do with the relations between the different quantum Cauchy surfaces, which by construction must base on the algebra of the quantum constraints. From the physical point of view, the quantum constraints should generate the deformations of the quantum Cauchy surfaces. Indeed, since the rigging map by construction is invariant under the transformations generated by the quantum constraints, we have 
\begin{eqnarray}
\label{deform 1}
\hat{\mathbb P}\, e^{i\,\hat C^{\text N}_i\cdot\hat{N}^i_{\text N}} \,\hat{\Pi}_t=\hat{\mathbb P}\hat{\Pi}_t=\hat{I},
\end{eqnarray}
where the lapse and shift operators $\hat{N}^i_{\text N}$ could be arbitrary kinematic operators, as long as it is arranged to the right of $\hat C^{\text N}_i$. Therefore, these transformations indeed deforms $\hat{\Pi}_t$ into another quantum Cauchy surface, so we may write 
\begin{eqnarray}
\label{deform 2}
e^{i\,\hat C^{\text N}_i\cdot\hat{N}^i_{\text N}}\, \cdot\,\hat{\Pi}_t \equiv \hat{\Pi}_{t;\hat{N}}.
\end{eqnarray}
One may then define a one parameter family of quantum Cauchy surfaces $\hat{\Pi}_{\tau_0;\hat{N}(\tau)}$ deformed from the $\hat{\Pi}_{\tau_0}$ (with $\hat{N}_{\text N}^i(\tau_0)=0$). This family give a foliation to the set of quantum spacetimes $\mathbb D_{\tau_0}$.
In terms of such deformations we may write down a sufficient condition for the domain stability, which is 
\begin{eqnarray}
\label{suff}
Image\big[e^{i\,\hat C^{\text N}_i\cdot\hat{\bar N}^i_{\text N}(\tau)}|_{\mathbb{S}^{\tau_0}_\star}\big]=\mathbb{S}^{\tau}_\star 
\end{eqnarray}
for some $\hat{\bar N}_{\text N}^i(\tau)$ and over some finite interval $\tau_0\leq \tau \leq \tau''_1$. The sufficient condition is thus the existence of the specific deformations that translate the clock fields' values according to ${\phi}^{\text N}_i=\bar{\phi}^{\text N}_i(\tau)$. When this condition is met, the domain stability would be granted in the interval with $\mathbb D_{\tau}=\mathbb D_{\tau_0}=\mathbb D$. In many homogeneous cosmological models with only one decoupled clock variable, the requirement \eqref{suff} can be achieved easily through setting the lapse operator inverse to the clock momentum operator, which leads to a translation generator for the clock variable. In a generalized case, the validity of \eqref{suff} depends greatly on the details of the theory, especially on the explicit quantization of the constraints and the resulted algebra in the quantum level.  

Finally, whenever the two stability conditions are established in the respective intervals, we expect a Schr\"odinger theory defined in the overlapping interval $\tau_0\leq \tau \leq \tau_1$ with $min\{\tau'_1,\tau''_1\}\equiv \tau_1$.

\subsection{ Path Integral and Schr\"odinger Transition Amplitudes}

By construction, the matrix elements of the rigging map provide the full information of the Dirac theory. Thus we should be able to formulate our construction in terms of these matrix elements. Specifically, given a background $\bar{\phi}^{\text N}_i(\tau)$, we would like to extrapolate the Schr\"odinger theories using only the transition amplitudes among the associated sectors of $\{\mathbb{S}^{{\tau}}_\star\}$. As mentioned, these elements are expected to be the transition amplitudes given by a path integral theory \cite{foam1}\cite{foam2}\cite{path1}\cite{path2}, so the formulation may also provide an interpretation to the path integral amplitudes in terms of the Schr\"odinger picture. In the following, we describe a procedure of using the matrix elements of the associated sectors to obtain the Schr\"odinger theories under the given background.

Starting from the ${\mathbb S}_{\star}^{\tau}$ specified by the background $\bar{\phi}^{\text N}_i(\tau)$, our first step is to identify the rigging map kernels $\{Ker[\hat{\mathbb{P}}|_{{\mathbb S}_{\star}^{\tau}}]\}$ of various $\tau$ values. Recall that these kernels are specified by \eqref{zero kernel}, which states that the kernels are given by the members in $\{\mathbb{S}^{{\tau}}_\star\}$ with zero self-transition amplitudes. Therefore, the matrix elements $\{\bra{\bar{\phi}^{\text N}_i(\tau),\psi'^{K}_i}\hat{\mathbb P}\ket{\bar{\phi}^{\text N}_i(\tau),\psi^{K}_i}\}$ in $\mathbb{S}^{{\tau}}_\star$ can determine $Ker[\hat{\mathbb{P}}|_{{\mathbb S}_{\star}^{\tau}}]$ for each $\tau$.

After the kernels are determined, one may choose $\mathbb{S}^{{\tau}}= Span\{\ket{\bar{\phi}^{\text N}_i(\tau),{\bold X}_{{\bold N}}}\}$ that is complement to the kernel at each $\tau$ value, over a finite interval $\tau_0\leq \tau \leq \tau'_1$ of our concern. From the argument in the previous section, we expect this to be achievable quite generally. The complete set $\{\hat{\bold X}_{{\bold N}}\}$ in $\mathbb{S}^{{\tau}}$ then induces a complete set of Dirac observables in $\mathbb D_{\tau}$, with an orthonormal eigenbasis given by \eqref{eigenstate 1}. Denoting the set of eigenvalues for $\hat{\bold X}_{{\bold N}}$ as $\{{\bold X}^{{(I)}}_{{\bold N}}\}$, we can write down the inner product between any two of the basis members in $\tau_0\leq \tau \leq \tau'_1$, as
\begin{eqnarray}
\label{trans amp0}
\big({{\bold X}^{{(I)}}_{{\bold N}}}{ ; \, \hat{\Lambda}_{\mathbb S^{\tau'}}}\big|{{\bold X}^{{(J)}}_{{\bold N}}}{ ; \, \hat{\Lambda}_{\mathbb S^{\tau}}}\big)= {(\bold\Lambda^{\!-\!1}_{\mathbb S^{{\tau}'}})^*}^K_I \,{(\bold\Lambda_{\mathbb S^{\tau}}^{\!-\!1})}^L_J\,\, \bra{\bar{\phi}^{\text N}_i(\tau'),{\bold X}^{{(K)}}_{{\bold N}}} \,\hat{\mathbb P} \,\ket{\bar{\phi}^{\text N}_i(\tau),{\bold X}^{{(L)}}_{{\bold N}}}.
\end{eqnarray}
Setting $\tau=\tau'$, the orthonormal condition for the eigenbasis leads to \eqref{Lambda1} in the form 
\begin{eqnarray}
\label{Lambda2}
 {({\bold\Lambda}^{\!-\!1}_{\mathbb S^{\tau}})^*}_L^J\,\,{({\bold\Lambda}^{\!-\!1}_{\mathbb S^{\tau}})}_M^K \,\,\bra{\bar{\phi}^{\text N}_i(\tau),{\bold X}^{{(J)}}_{{\bold N}}}\hat{\mathbb{P}}\ket{\bar{\phi}^{\text N}_i(\tau),{\bold X}^{{(K)}}_{{\bold N}}}
=\delta_{L,M}.
\end{eqnarray}
One may then solve the above for $\hat{\bold\Lambda}_{\mathbb S^{\tau}}={\bold\Lambda}_{\mathbb S^{\tau}}(\hat{\bold{X}}_{{\bold N}}, \hat{\bold{P}}_{{\bold N}})={\bold\Lambda}(\hat{\phi}^{\text N}_i,\hat{\bold{X}}_{{\bold N}}, \hat{\bold{P}}_{{\bold N}})$ using the matrix elements $\{\bra{\bar{\phi}^{\text N}_i(\tau),{\bold X}^{{(J)}}_{{\bold N}}}\hat{\mathbb{P}}\ket{\bar{\phi}^{\text N}_i(\tau),{\bold X}^{{(K)}}_{{\bold N}}}\}$.
Finally, with the $\hat{\bold\Lambda}_{\mathbb S^{\tau}}$ determined, we may use the matrix elements $\{\bra{\bar{\phi}^{\text N}_i(\tau'),{\bold X}^{{(J)}}_{{\bold N}}}\hat{\mathbb{P}}\ket{\bar{\phi}^{\text N}_i(\tau),{\bold X}^{{(K)}}_{{\bold N}}}\}$ to evaluate values of the inner products \eqref{trans amp0} with $\tau'\neq \tau$, and if one finds 
\begin{eqnarray}
\label{trans amp2}
\sum_{I}\big|\big({{\bold X}^{{(I)}}_{{\bold N}}}{ ; \, \hat{\Lambda}_{\mathbb S^{\tau'}}}\big|{{\bold X}^{{(J)}}_{{\bold N}}}{ ; \, \hat{\Lambda}_{\mathbb S^{\tau}}}\big)\big|^2=1\,
\end{eqnarray}
for $\tau_0\leq \tau \leq \tau''_1$ then we have $\mathbb D_{\tau'}=\mathbb D_{\tau}= \mathbb D$ in the interval. In this case, the matrix \eqref{trans amp0} gives the unitary evolution operator for the wave function 
\begin{eqnarray}
\label{wave function 5}
\bold{\Psi}_{\mathbb D\,(\tau)}\big[{\bold X}^{{(I)}}_{{\bold N}}{ ; \, \hat{\Lambda}_{\mathbb S^{\tau}}}\big]\equiv \big({{\bold X}^{{(I)}}_{{\bold N}}}{ ; \, \hat{\Lambda}_{\mathbb S^{\tau}}}\big|\Psi_{\mathbb D}\big)\,.
\end{eqnarray}
If \eqref{trans amp2} does not hold for any interval, then there is no domain stability and no Schr\"odinger theory is obtained.
Thus, the transition amplitudes between $\mathbb S^{\tau}$ and $\mathbb S^{\tau'}$ detect the domain stability, and in the presence of the stability they also govern the evolution of the wave functions. 

In the above we have shown a procedure of using the path integral transition amplitudes $\{ \bra{\bar{\phi}^{\text N}_i(\tau'),\psi'^{K}_i}\hat{\mathbb P}\ket{\bar{\phi}^{\text N}_i(\tau),\psi^{K}_i}\}$ to extract the Schr\"odinger theories with the given background.

We now address the possible transformations relating one quantum foliation to another. Suppose we have two Schr\"odinger theories with their own sets of background fields $\bar{\phi}^{\text N}_i=\bar{\phi}^{\text N}_i(t)$ and ${}^{*}\bar{\phi}^{\text N}_i={}^{*}\bar{\phi}^{\text N}_i(\tau)$, respectively defined in the domains $\mathbb D$ and ${}^{*}\mathbb D$. If ${}^{*}\mathbb D \subset \mathbb D$, then every wave function $\bold{\Psi}_{(\tau)}\big[{}^{*}{\bold X}_{{\bold N}}{ ; \, \hat{\Lambda}_{\mathbb S^{\tau}}}\big]$ describing a state $\Psi\in {}^{*}\mathbb D \subset \mathbb D$ can be transformed into 
$\bold{\Psi}_{(t)}\big[{\bold X}_{{\bold N}}{ ; \, \hat{\Lambda}_{\mathbb S^{t}}}\big]$ through the matrix  
\begin{eqnarray}
\label{trans amp1}
\big({{\bold X}^{{(I)}}_{{\bold N}}}{ ; \, \hat{\Lambda}_{\mathbb S^{t}}}\big|{{}^{*}{\bold X}^{{(J)}}_{{\bold N}}}{ ; \, \hat{\Lambda}_{\mathbb S^{\tau}}}\big)= {{(\bold\Lambda^{\!-\!1}_{\mathbb S^{t}}})^*}^K_I \,{(\bold\Lambda_{\mathbb S^{\tau}}^{\!-\!1})}^L_J\,\, \bra{\bar{\phi}^{\text N}_i(t),{\bold X}^{{(K)}}_{{\bold N}}} \,\hat{\mathbb P} \,\ket{{}^{*}\bar{\phi}^{\text N}_i(\tau),{}^{*}{\bold X}^{{(L)}}_{{\bold N}}}.
\end{eqnarray}
Physically, this transformation is to switch from the foliation in $\tau$ to that in $t$ for the quantum spacetimes in ${}^{*}\mathbb D$.  Also, it is given by the path integral transition amplitudes between $\mathbb S^{\tau}$ and $\mathbb S^{t}$. Moreover, a state $\Psi' \in (\mathbb D - {}^{*}{\mathbb D})$ has a unitary Schr\"odinger representation only under the background $\bar{\phi}^{\text N}_i(t)$, but not under ${}^{*}\bar{\phi}^{\text N}_i(\tau)$. The reason is clear -- while $\{(\!\hat{\Lambda}_{\mathbb S^{\tau}}^{\!-\!1}{}^{*}\!\hat{X}^{{{\mathbb{S}^{{\tau}}}}}_{{\bold N}}\!\hat{\Lambda}_{\mathbb S^{\tau}})(\tau)\}$ is a complete set of observables in ${}^{*}\mathbb D$,  it is not a complete set in $\mathbb D$. In the special cases of $\mathbb D= {}^{*}\mathbb D$, the two Schr\"odinger theories are dual to each other, as the descriptions from the two different foliations of the same set of quantum spacetimes. If $(\mathbb D - {}^{*}{\mathbb D})\neq \emptyset$, the theory $(\mathbb D, \{\hat{\Pi}_{t}\}_{t_0 \leq t \leq t_1})$ is more global since it applies to a broader range of measurements. 

We've mentioned the theories of quantum gravity \cite{reduced2}\cite{reduced3} with an absolute notion of time defined under a privileged physical background $\bar{\phi}^{\text N}_{i\, {Abs}}(t_{ {Abs}})$, such that the universe is described with a fundamental Schr\"odinger theory. We also raised the question about whether such a theory could account for the possibly detectable quantum behavior of the background fields ${\phi}^{\text N}_{i\, {Abs}}$.
Here we offer a possible scenario in which the answer is positive. The physical subspace that can support a fundamental Schr\"odinger theory may turn out to be an exceptional domain $\mathbb D_{ {Abs}}$. In such consideration, this domain would be regarded as the true physical Hilbert space $\mathbb H_{{phys}}\equiv \mathbb D_{ {Abs}}$. In the fundamental Schr\"odinger theory $(\mathbb D_{ {Abs}}, \{\hat{\Pi}_{t_{ {Abs}}}\}_{-\infty \leq t_{ {Abs}} \leq \infty})$, the fields ${\phi}^{\text N}_{i\, {Abs}}$ are without quantum fluctuations. Nevertheless, there can be a sub domain ${}^{*}{\mathbb D}\subset \mathbb D_{ {Abs}}$ that provides an effective Schr\"odinger theory $({}^{*}{\mathbb D}, \{\hat{\Pi}_{\tau}\}_{\tau_0 \leq \tau\leq \tau_1})$ in a different foliation, in which ${\phi}^{\text N}_{i\, {Abs}}$ are dynamical quantum fields.

\section{One Dimensional Model}

Here we apply the quantum Cauchy surfaces to a one dimensional model with a single constraint. The classical system of our model has a kinematic phase space coordinatizd by three canonical conjugate pairs $ \{X_j, P_j\}_{j=A,B,C}$. The non-trivial Poisson brackets are given by $\{X_j, P_j\}= 1 $, and the system is governed by the scalar constraint $H\equiv P_A^2/m_A+ m_A \omega^2 X_A^2 - P_B^2/m_B + P_C^2/m_C$.  

\subsection{ Refined Algebraic Quantization and Timeless Physical Hilbert Space $\mathbb H$}

To obtain the Dirac theory for the system, we first quantize the system through the refined algebraic quantization procedure.
The canonical quantization of the kinematic phase space leads to the quantized constraint operator  $\hat{H}\equiv \hat{P}_A^2/m_A +  m_A \omega^2\hat{X}_A^2 - \hat{P}_B^2/m_B + \hat{P}_C^2/m_C$, which acts on the kinematic Hilbert space $\mathbb{K}\equiv Span\{\ket{E_n^A ,{P}_B,{P}_C }\}$, where $E_n^A$ denotes the harmonic oscillator's energy levels of the subsystem $A$. In this case, the physical Hilbert space $\mathbb{H}\subset \mathbb{K}^*$ can be constructed through the standard group averaging method, which gives the rigging map $\hat{\mathbb{P}}: {\mathbb{K}} \to {\mathbb{H}}$ as:
\begin{equation}
\begin{split}
\label{riggin map}
\hat{\mathbb{P}}\equiv \int_{-\infty}^{\infty} d\lambda e^{i\lambda\hat{H}}
= \delta\left( \hat{P}_A^2/m_A + m_A \omega^2\hat{X}_A^2 - \hat{P}_B^2/m_B + \hat{P}_C^2/m_C\right).
\end{split}
\end{equation}
The rigging map then equips $\mathbb{H}$ with a Hermitian inner product. As mentioned, the inner product between any two physical states $|\Psi_1)\equiv \hat{\mathbb{P}}\ket{\psi_1}$ and $|\Psi_2)\equiv \hat{\mathbb{P}}\ket{\psi_2}$ is given by 
\begin{equation}
\label{inner}
\begin{split}
(\Psi_1|\Psi_2)\equiv \bra{\psi_1}\hat{\mathbb{P}} \ket{\psi_2},
\end{split}
\end{equation}
where $\braket{\cdot | \cdot}$ and $(\cdot|\cdot)$ denote respectively the inner products in $\mathbb{K}$ and $\mathbb{H}$. 

Denote the energy levels of the system $A$ as $E_n^A= \hbar\omega(2n+1)$. We also introduce the following useful notations
\begin{eqnarray}
\label{notation1}
 \eta_B\equiv \pm 1 \,\,; \,\,\,\,|P_B^{\star}|(n,P_C)&\equiv&\sqrt{m_B(E_n^A+{P}_C^2/m_C)}\,\,; \,\,\,P_B^{\star}\equiv \eta_B |P_B^{\star}|\,\,; \nonumber\\
\eta_C\equiv \pm 1 \,\,; \,\,\,\,|P_C^{\star}|(n,P_B)&\equiv&\sqrt{m_C|{P}_B^2/m_B-E_n^A|}\,\,; \,\,\,P_C^{\star}\equiv \eta_C |P_C^{\star}|.
\end{eqnarray}
 We have 
\begin{eqnarray} 
\label{basis 1}
\hat{\mathbb P}\,\, \lim_{\epsilon\to 0}\,\, \int_{{P}_B^\star-\epsilon}^{{P}_B^\star+\epsilon }dP'_B\,\,\ket{E_n^A,P'_B ,{P}_C }&=&\frac{1}{{|{P}_B^\star|}/2m_B}\ket{E_n^A\,,\,P_B^{\star}\, ,\,{P}_C },\nonumber\\
\text{and}\,\,\,\,\hat{\mathbb P}\,\, \lim_{\epsilon\to 0}\,\, \int_{{P}_C^\star-\epsilon}^{{P}_C^\star+\epsilon }dP'_C\,\,\ket{E_n^A,P_B ,{P}'_C }_{{P}_C^{\star2}\geq 0}&=&\frac{1}{{|{P}_C^\star|}/2m_C}\ket{E_n^A\,,\,P_B\, ,\,{P}_C^{\star} }_{{P}_C^{\star2}\geq 0}.\nonumber\\
\end{eqnarray}
Using the above and the inner product defined in \ref{inner}, we find $\mathbb{H}$ to be spanned by either of the  ($\delta$- normalized) orthonormal basis
\begin{eqnarray}
\label{basis 2}
\bigg\{ \bigg| E_n^A, \eta_B ,P_C\bigg)&\equiv&\frac{1}{\sqrt{{|{P}_B^\star|}/2m_B}}\ket{E_n^A\,,\,P_B^{\star}\, ,\,{P}_C }\bigg\}\nonumber \\
\text{and}\,\,\,\bigg\{ \bigg| E_n^A, P_B  ,\eta_C\bigg)_{{P}_C^{\star2}\geq 0}&\equiv&\frac{1}{\sqrt{{|{P}_C^\star|}/2m_C}}\ket{E_n^A\,,\,P_B\, ,\,{P}^{\star}_C }_{{P}_C^{\star2}\geq 0}\bigg\}\nonumber\\
\end{eqnarray}
 satisfying 
\begin{eqnarray}
\label{basis 3}
\bigg(E_m^A,\eta'_B, P'_C \bigg| E_n^A,\eta_B,P_C\bigg)=\delta_{m,n} \delta_{\eta'_B,\eta_B} \delta(P'_C-P_C),\\
\text{and}\,\,\, \bigg(E_m^A,P'_B  ,\eta'_C \bigg| E_n^A,P_B  ,\eta_C\bigg)_{{P}_C^{\star2},{P}_C^{'\star2}\geq 0}=\delta_{m,n} \delta_{\eta'_C,\eta_C} \delta(P'_B-P_B).
\end{eqnarray}
A typical physical state in $\mathbb H$ thus takes the forms
\begin{eqnarray}
\label{heisswave}
\big|\Psi\big)
&=&
 \sum_{n}\int_{-\infty}^{\infty} d P_C\,\, \Psi_{H_{{}^{;C}}}(E_n^A,\eta_B,P_C) \big|E_n^A,\eta_B,P_C\big),\nonumber\\
&=&
\sum_{n} \int_{{P}_C^{\star2}\geq 0} d P_B\,\, \Psi_{H_{{}^{;B}}}(E_n^A,P_B  ,\eta_C) \big|E_n^A,P_B  ,\eta_C\big).
\end{eqnarray}
where $  \Psi_{H_{{}^{;C}}} =\sqrt{{|{P}_C^\star|}/2m_C}/\sqrt{{|{P}_B^\star|}/2m_B}\,\, \Psi_{H_{{}^{;B}}}$ denotes the timeless and normalized Heissenberg wave function. Finally, all the self-adjoint operators in $\mathbb K$ commuting with $\hat H$ become Dirac observables in $\mathbb H$.

Having completed the refine algebraic quantization, we now apply the quantum Cauchy surfaces to the timeless theory to obtain the Schr\"odinger theory under a specified background. Particularly, we will follow the procedure using the relevant transition amplitudes, which is described in Sec.II.D.

{{{{{{{{{{{{{{{{{{{{{{{{{{{{{{{{{{{{{{{{{{{{{{{{{{{{{{{{{{{{{{{{
\subsection{ Quantum Cauchy Surfaces with $\hat{\phi} \equiv \hat{X}_B$}

In this section we look for the Schr\"odinger theory in which $X_B$ appears to be a classical monotonic background specified as $\bar{X}_B(t)\equiv \bar{\phi}(t)= t $. According to the background, the relevant eigenspace for each $t$ is given by $\mathbb S^{t}_\star=Span\big\{\ket{E_n^A,\bar{X}^{\pm}_{B}(t), P_C}\big\}$ spanned by the basis members of the infinitely wide wave packets, each with the specified value in $X_B$ and a definite sign in $P_B$. The basis is thus defined by
\begin{equation}
\begin{split}
\label{packet}
\ket{\bar{X}^{\pm}_B(t)}\equiv \pm \int_0^{\pm\infty} dP'_B \,e^{-iP'_B \bar{X}_B(t)/\hbar} \ket{P'_B}.
\end{split}
\end{equation}

First, we calculate the transition amplitudes between the members in $\mathbb S^{t}_\star$ and $\mathbb S^{t'}_\star$. They can be obtained easily as
\begin{eqnarray} 
\label{trans B}
\bra{E_{n'}^A\,,\,\bar{X}^{{\eta}'_B}_{B}(t')\,,\,{P}'_C} \hat{\mathbb P}\ket{E_{n}^A\,,\,\bar{X}^{{\eta}_B}_{B}(t)\,,\,{P}_C}
=\frac{e^{i{P}_B^\star\, (t-t')/\hbar}}{{|{P}_B^\star|}/2m_B} \,\,\delta_{{\eta}'_B, {\eta}_B}\,\, \delta_{n',\,n}\,\, \delta ({P}'_C-{P}_C).
\end{eqnarray}
By setting $t'=t$, we see that the equation \eqref{zero kernel} in this case has only the zero vector solution for any $t$ so the kernel is trivial. Thus we have the quantum Cauchy surface $\hat{\Pi}_t$ with $Image\{\hat{\Pi}_t|_{\mathbb{D}_t}\}=\mathbb S^{t}=\mathbb S^{t}_\star=\big\{Span\big\{\ket{E_n^A,\bar{X}^{\pm}_{B}(t), P_C}\big\}$ for each value of $t$. Then, we may set $\{{\bold X}^{{{\mathbb{S}^{{t}}}}}_{{\bold N}}\}\equiv \{{\bold X}_{{\bold N}}\}\equiv\{ \hat{X_A} , \hat{\eta}_B  , \hat{X}_C\}$, and $\{{\bold P}^{{{\mathbb{S}^{{t}}}}}_{{\bold N}}\}\equiv \{{\bold P}_{{\bold N}}\}\equiv\{ \hat{P_A} , \hat{P}_{{{\eta}_B}}  , \hat{P}_C\}$, where $\hat{P}_{{{\eta}_B}}$ is conjugate to $\hat{\eta}_B$.

Next, we look into the transition amplitudes between the two sets of eigenbasis at $t$ and $t'$, given by
\begin{eqnarray} 
\label{trans B2}
\bra{\bar{\phi}(t'),{\bold P}'_{{\bold N}}}\hat{\mathbb{P}}\ket{\bar{\phi}(t),{\bold P}_{{\bold N}}}\equiv\bra{E_{n'}^A\,,\,\bar{X}^{{\eta}'_B}_{B}(t')\,,\,{P}'_C} \hat{\mathbb P}\ket{E_{n}^A\,,\,\bar{X}^{{\eta}_B}_{B}(t)\,,\,{P}_C}
\end{eqnarray}
By setting $t=t'$ in the above, we solve the equation \eqref{Lambda2} and find an obvious solution 
\begin{eqnarray} 
\label{Lambda B}
\hat{\bold{\Lambda}}_{\mathbb S^{t}}\equiv \frac{1}{\sqrt{{|\hat{P}_B^\star|}/2m_B}}\,\,;\,\,\hat{P}_B^\star\equiv {P}_B^\star(\hat{\eta}_B,\hat{X_A}, \hat{P_A}, \hat{P_C}).
\end{eqnarray}
Then, a complete set of Dirac observables for $\mathbb D_t$ may be given by $\{\big( \hat{\Lambda}_t^{\!-\!1} { X}^{{{\mathbb{S}^{{t}}}}}_{{\bold N}}\hat{\Lambda}_t\big)(t)\big\}$ or $\{\big( \hat{\Lambda}_t^{\!-\!1} { P}^{{{\mathbb{S}^{{t}}}}}_{{\bold N}}\hat{\Lambda}_t\big)(t)\big\}$, which satisfies
\begin{eqnarray}
\label{observables B}
\big( \hat{\Lambda}_t^{\!-\!1}\, ({X}^{{{\mathbb{S}^{{t}}}}}_{{\bold N}}\,, \,{P}^{{{\mathbb{S}^{{t}}}}}_{{\bold N}})\,\hat{\Lambda}_t\big)(t)\bigg|_{\mathbb D_t}
&\equiv&\,\, \hat{\mathbb P}\,\,\sqrt{{|\hat{P}_B^\star|}/2m_B}\,\,(\hat{\bold X}_{{\bold N}}\,,\,\hat{\bold P}_{{\bold N}})\,\,\frac{1}{\sqrt{{|\hat{P}_B^\star|}/2m_B}}\,\, \hat{\Pi}_t\,.
\end{eqnarray} 

Next, inserting the values of \eqref{trans B2} and the given $\hat{\Lambda}_t$ into \eqref{trans amp0}, we find that \eqref{trans amp2} is satisfied and thus $\mathbb D_t = \mathbb D$ for all $t$. Therefore, we have now identified a Schr\"odinger theory $(\mathbb D, \{\hat{\Pi}_{t}\}_{-\infty\leq t \leq \infty})$ under the background $\bar{X}_B(t)=t$, with the wave functions of the form $\bold{\Psi}_{\mathbb D\,(t)}\big[{\bold X}_{{\bold N}}{ ; \, \hat{\Lambda}_{\mathbb S^{t}}}\big]$ as defined in \eqref{wave function 5}. Finally, using a more convenient complete set $ \{\tilde{\bold X}_{\bold N}\}\equiv\{ \hat{E_n^A} , \hat{\eta}_B  , \hat{P}_C\}$ we can calculate the evolution matrix for the wave functions through \eqref{trans amp2}, which is just
\begin{eqnarray}
\big(\tilde{\bold X}'_{\bold N}{ ; \, \hat{\Lambda}_{\mathbb S^{t'}}}\big|\tilde{\bold X}_{\bold N}{ ; \, \hat{\Lambda}_{\mathbb S^{t}}}\big)
= \,\,e^{i{P}_B^\star\, (t'-t)/\hbar} \,\,\delta_{{\eta}'_B, {\eta}_B}\,\, \delta_{n',\,n}\, \delta ({P}'_C-{P}_C).
\end{eqnarray}
One can now read off the effective Hamiltonian governing the wave function $\bold{\Psi}_{\mathbb D\,(t)}\big[{\bold X}_{{\bold N}}{ ; \, \hat{\Lambda}_{\mathbb S^{t}}}\big]$, which is
\begin{eqnarray}
\label{Hamiltonian1}
\hat{\bold H}_B= {\bold H}_B(\hat{\bold X}_{{\bold N}}\,,\,\hat{\bold P}_{{\bold N}})\equiv \hat{P}_B^\star = \hat{\eta}_B\sqrt{m_B(\hat{E}_n^A+\hat{P}_C^2/m_C)}.
\end{eqnarray}
Observe that the theory is identical to the simpliest theory obtained through the reduced phase space quantization using $X_B$ as the clock.
Lastly, we can check that $\mathbb D= \mathbb H$, so the Schr\"odinger theory $(\mathbb D, \{\hat{\Pi}_{t}\}_{-\infty\leq t \leq \infty})$ is globally defined in the full physical Hilbert space.

As a final remark we make an observation on the state of the background. Had we chosen $\ket{\bar{X}_{B}(t)}$ as the state of the background, we would have $\mathbb S^{'t}_{\star}=\mathbb S^{'t}= Span\big\{\ket{E_n^A,\bar{X}_{B}(t), P_C}\big\}$. Correspondingly, we would find $\mathbb{D}'_t=Span\big\{e^{-i{P_B^\star\,\bar{X}_B(t)/\hbar}}\big| E_n^A, +\, ,P_C\big) + e^{i{P_B^\star\,\bar{X}_B(t)/\hbar}}\big| E_n^A, - \,,P_C\big)\big\}$. One can then immediately see that $\mathbb{D}'_{t'} \neq \mathbb{D}'_{t}$ whenever $t'\neq t$, and thus there is no unitary evolution in any range of $t$.

{{{{{{{{{{{{{{{{{{{{{{{{{{{{{{{{{{{{{{{{{{{{{{{{{{{{{{{{{{{{{{{{
\subsection{ Quantum Cauchy Surfaces with ${}^{*}\!\hat{\phi}= \hat{X}_A $}

Now we look for the Schr\"odinger theory in which $X_A$ appear to be a background with the assigned value $\bar{X}_A(\tau)\equiv{}^{*}\!\hat{\phi}(\tau)=\tau$. According to the background, the relevant eigenspace for each $\tau$ is given by $\mathbb S^{\tau}_\star=Span\big\{\ket{\bar{X}^{\pm}_{A}(\tau),P_B,P_C}\big\}$, where $\ket{\bar{X}^{\pm}_{A}}$ is analogously given by \eqref{packet}.

First, we calculate the transition amplitudes between the members in $\mathbb S^{\tau}_\star$ and $\mathbb S^{\tau'}_\star$. With $\phi_n(\bar{X}^\pm_A) \equiv\braket{E_n^A|\bar{X}^\pm_A}$, they are given by
\begin{eqnarray} 
\label{trans A}
\bra{\bar{X}^{{\eta}'_A}_{A}(\tau')\,,\,{P}'_B\,,\,{P}'_C\,} \,\hat{\mathbb P}\,\ket{\,\bar{X}^{{\eta}_A}_{A}(\tau)\,,\,{P}_B\,,\,{P}_C}
&=&\sum_{n',\,n}\,\,\delta(0)^2\,\,\frac{\phi^*_{n}(\bar{X}^{{\eta}'_A}_{A}(\tau'))\,\,\phi_{n'}(\bar{X}^{{\eta}_A}_{A}(\tau))  }{{|{P}_B|}/2m_B} \nonumber\\
&\times&
\delta_{|P_B|, |\,P_B^\star|}\,\,\delta_{|P_B|, \,|{P}_B^{'\star}|}\,\,\delta_{P_B, P'_B}\,\,\delta(P_C-P'_C),\nonumber\\
\end{eqnarray}
where the $\delta(0)^2$ can be tamed by a smearing over both ${P}'_B$ and ${P}_B$.

By setting $\tau'=\tau$ in the above, we again use \eqref{zero kernel} to look for the kernel. This time, we have a non-trivial kernel given by
\begin{eqnarray} 
\label{kernel A}
Ker[\hat{\mathbb{P}}|_{{\mathbb S}_{\star}^{\tau}}]&=& Span\big\{\ket{\bar{X}^{\pm}_{A}(\tau),P_B,P_C}_{P_B \neq P_B^\star(n,P_C) \,\forall n}\,\big\}\nonumber\\
&&\oplus\, Span\big\{\ket{\bar{X}^{+}_{A}(\tau),P_B,P_C}- \kappa_{{P_B,P_C}}\ket{\bar{X}^{-}_{A}(\tau),P_B,P_C}\big\}. 
\end{eqnarray}
The first component of the kernel says that the physical spectrum of $P_B$ corresponding to each $P_C$ should be discretized according to the values of $E_n^A$; the second component indicates the degeneracy between ${\eta}_A=\pm 1$ for the physical states due to the periodicity of the oscillator. This gives us a natural choice for a compliment space, given by $Image\{\hat{\Pi}_{\tau}|_{\mathbb{D}_\tau}\}={\mathbb S}^{\tau}\equiv Span\{\lim_{\epsilon\to 0}\,\epsilon \ket{\,\bar{X}^{+}_{A}(\tau)\,,\,{P}_B^\star\,,\,{P}_C}\}$ with
\begin{eqnarray}
\label{basis A}
\lim_{\epsilon\to 0}\epsilon \ket{\,\bar{X}^{+}_{A}(\tau)\,,\,{P}_B^\star\,,\,{P}_C} \equiv
\lim_{\epsilon\to 0}\,\, \int_{\,{P}_B^\star-\epsilon/2}^{{P}_B^\star+\epsilon/2} dP'_B \,\,\ket{\,\bar{X}^{{\eta}_A}_{A}(\tau)\,,\,P'_B\,,\,{P}_C}.
\end{eqnarray}
One obvious complete set of self-adjoint operators in ${\mathbb S}^{\tau}$ is given by $\{\hat{\bold P}^{{{\mathbb{S}^{{\tau}}}}}_{{\bold N}}\}\equiv \{{}^{*}\hat{\bold P}_{{\bold N}}\}\equiv\{\hat{P}_B  , \hat{P}_C\}$. Tailored to the partially discrete spectrum of $(P_B^\star, P_C)$ in ${\mathbb S}^{\tau}$, we can construct two commuting ``difference" operators $\hat{X}^\centerdot_B$ and $\hat{X}^\centerdot_C$, such that $\{{}^{*}\hat{\bold X}_{{\bold N}}\}\equiv\{\hat{X}^\centerdot_B,\hat{X}^\centerdot_C\}$ is another complete set for ${\mathbb S}^{\tau}$ conjugate to the set$\{{}^{*}\hat{\bold P}_{{\bold N}}\}$. Also, since the spectrum gap of $(P_B^\star, P_C)$ is of $\mathcal{O}(\hbar)$, we naturally have $\lim_{\hbar\to 0}(\hat{X}^\centerdot_B,\hat{X}^\centerdot_C)=(\hat{X}_B,\hat{X}_C)$.

Next, we look into the transition amplitudes between the two sets of eigenbasis at $\tau$ and $\tau'$, given by
\begin{eqnarray} 
\label{trans B2}
\bra{{}^{*}\!\bar{\phi}(\tau'),{}^{*}{\bold P}'_{{\bold N}}}\hat{\mathbb{P}}\ket{{}^{*}\!\bar{\phi}(\tau),{}^{*}{\bold P}_{{\bold N}}}\equiv \lim_{\epsilon\to 0}\epsilon^2\bra{\,\bar{X}^{+}_{A}(\tau')\,,\,{{P}'_B}^{\star}\,,\,{P}'_C} \hat{\mathbb P}\ket{\,\bar{X}^{+}_{A}(\tau)\,,\,{P}_B^\star\,,\,{P}_C}
\end{eqnarray}
whose values are given by \eqref{trans A} by noting that $\lim_{\epsilon\to 0}\epsilon^2 \delta(0)^2=1$. Setting $\tau=\tau'$ we solve \eqref{Lambda2} and find a solution to be
\begin{eqnarray} 
\label{Lambda A}
\hat{\bold{\Lambda}}_{\mathbb S^{\tau}}\equiv \lim_{\epsilon\to 0}\,\frac{|\phi_{\hat n}(\bar{X}^{+}_{A}(\tau))|}{ \epsilon\,\,\sqrt{{|\hat{P}_B|}/2m_B}}\,;\,\, \hat{n}\equiv n(\hat{P}_B, \hat{P}_C). 
\end{eqnarray}
Then a complete set of Dirac observables for $\mathbb D_\tau$ may be given by $\{\big( \hat{\Lambda}_\tau^{\!-\!1} { X}^{{{\mathbb{S}^{{\tau}}}}}_{{\bold N}}\hat{\Lambda}_\tau\big)(\tau)\big\}$ or $\{\big( \hat{\Lambda}_\tau^{\!-\!1} { P}^{{{\mathbb{S}^{{\tau}}}}}_{{\bold N}}\hat{\Lambda}_\tau\big)(\tau)\big\}$ which satisfies
\begin{eqnarray}
\label{observables A}
\big( \hat{\Lambda}_\tau^{\!-\!1}\, ({X}^{{{\mathbb{S}^{{\tau}}}}}_{{\bold N}}\,,\,{P}^{{{\mathbb{S}^{{\tau}}}}}_{{\bold N}})\,\hat{\Lambda}_\tau\big)(\tau)\bigg|_{\mathbb D}
&\equiv&\,\, \hat{\mathbb P}\,\,\frac{ \,\,\sqrt{{|\hat{P}_B|}/2m_B}}{|\phi_{\hat n}(\bar{X}^{+}_{A}(\tau))|}\,\,({}^{*}\hat{\bold X}_{{\bold N}}\,,\,{}^{*}\hat{\bold P}_{{\bold N}})\,\,\frac{|\phi_{\hat n}(\bar{X}^{+}_{A}(\tau))|}{ \,\,\sqrt{{|\hat{P}_B|}/2m_B}}\,\, \hat{\Pi}_\tau\,.
\end{eqnarray} 
Note that, just as in general cases, the overall scaling like the $\epsilon$ in $\hat{\Lambda}_\tau$ does not appear in any of the corresponding Dirac observables.

 Next, inserting the values of \eqref{trans B2} and $\hat{\bold{\Lambda}}_{\mathbb S^{\tau}}$ into \eqref{trans amp0}, we find that \eqref{trans amp2} is again satisfied and thus $\mathbb D_\tau = {}^{*}\mathbb D$ for all $\tau$. Therefore, we have identified another Schr\"odinger theory $({}^{*}\mathbb D, \{\hat{\Pi}_{\tau}\}_{-\infty\leq \tau \leq \infty})$ under the background $\bar{X}_A(\tau)=\tau$, with the wave function $\bold{\Psi}_{{}^{*}\mathbb D\,(\tau)}\big[{}^{*}{\bold X}_{{\bold N}}{ ; \, \hat{\Lambda}_{\mathbb S^{\tau}}}\big]$ as defined in \eqref{wave function 5}. Finally, the evolution matrix for the wave functions defined in \eqref{trans amp2} is 
\begin{eqnarray}
\big({{}^{*}{\bold P}'_{{\bold N}}}{ ; \, \hat{\Lambda}_{\mathbb S^{\tau'}}}\big|{{}^{*}{\bold P}_{{\bold N}}}{ ; \, \hat{\Lambda}_{\mathbb S^{\tau}}}\big)
=  \,\,\frac{\phi^*_{ n}(\bar{X}^{+}_{A}(\tau'))\phi_{ n'}(\bar{X}^{+}_{A}(\tau))}{|\phi_{n}(\bar{X}^{+}_{A}(\tau'))||\phi_{n}(\bar{X}^{+}_{A}(\tau))|} \,\,\delta_{{P}_B,\,{P}'_B} \,\delta ({P}_C-{P}'_C).
\end{eqnarray}
To further evaluate this matrix, we introduce
\begin{eqnarray}
|\phi_n({X}^+_A)|\equiv|\phi|(n,{X}_A)\,; \,\,\phi_n({X}^+_A)= |\phi|(n,{X}_A) e^{-i S(n,{X}_A)/\hbar},
\end{eqnarray}
and by using the WKB approximation we can easily show that
\begin{eqnarray}
\label{WKB}
&&S(n,{X}_A)=\int_{X_0}^{{X}_A} dX_A \,\sqrt{m_A\left(E_n^A-m_A\omega^2X_A^2\right)} + \zeta(n,{X}_A)\,;\,\,\nonumber\\
&&|\phi|(n,{X}_A) \sim \mathcal{O}(\hbar^0)\,\,;\,\, \lim_{\hbar\to 0}\zeta(n,{X}_A)\big|_{E_n^A-m_A\omega^2X_A^2>0}=0.
\end{eqnarray}
Thus the above evolution matrix can be expressed as
\begin{eqnarray}
\exp\left[ -\frac{i}{\hbar}\int_{\tau}^{\tau'}  \,\partial_{X_A}\,S(n,\,{X}_A) \,\,d X_A  \right]\,\,\delta_{{P}_B,\,{P}'_B} \,\delta ({P}_C-{P}'_C)\,.
\end{eqnarray}
Finally, we recognize from the above the effective Hamiltonian, governing $\bold{\Psi}_{{}^{*}\mathbb D\,(\tau)}\big[{}^{*}{\bold X}_{{\bold N}}{ ; \, \hat{\Lambda}_{\mathbb S^{\tau}}}\big]$ under the background, is given by
\begin{eqnarray}
\label{Hamiltonian3}
\hat{\bold H}_{A\,(\tau)}={\bold H}_{A\,(\tau)}({}^{*}\hat{\bold X}_{{\bold N}},{}^{*}\hat{\bold P}_{{\bold N}})\equiv \partial_{\tau}\,S(\hat{n},\,\tau)=\sqrt{m_A\left(\hat{P}_B^{2}/m_B -\hat{P}_C^2/m_C-m_A \omega^2 \tau^2\right)} + \zeta(\hat{P}_B,\hat{P}_C, \tau).\nonumber\\
\end{eqnarray}
Referring to \eqref{WKB}, we see that the classical limits of the Shr\"odinger theory truly gives the classical reduced phase space theory using $X_A$ as the clock variable in the proper region satisfying $E_n^A-m_A\omega^2X_A^2>0$. Lastly, we can check that ${}^{*}\mathbb D= \mathbb H$, so the Schr\"odinger theory $({}^{*}\mathbb D, \{\hat{\Pi}_{\tau}\}_{-\infty\leq \tau \leq \infty})$ is globally defined in the full physical Hilbert space.

As a final remark, we comment on the proper background. Had we instead chosen $\ket{\bar{X}_{A}(\tau)}$ as the background state, we would have $ \mathbb{S}^{'\tau}_{\star}=\mathbb{S}^{'\tau}= Span\big\{\ket{\bar{X}_{A}(\tau),P_B^\star, P_C}\big\}$. In this case, because of the periodic nature of $A$, the domain stability ${{\mathbb{D}}}'_\tau={\mathbb{D}}=\mathbb H$ will still be satisfied. Also, the corresponding $\hat{{\Lambda}}'_\tau$ would be given by replacing the factor $\phi_{\hat{n}}({X}^+_A)$ in \eqref{Lambda A} with $\phi_{\hat n}({X}_A)$ ($\phi_{n}({X}_A)\equiv\braket{E_n^A|{X}_A}$). Since $\phi_n({X}_A)$ is a standing wave with zeros at the nodes, $\hat{{\Lambda}}_\tau^{'\!-\!1}$ diverges at these nodes. Unlike the overall scaling of the $\epsilon$, these infinities in $\hat{{\Lambda}}_\tau^{'\!-\!1}$ are operator divergences, because they depend on $\hat{n}=n(\hat{P}_B, \hat{P}_C)$. Therefore the quantum Cauchy surfaces with this alternative background do not lead to a well-defined Shr\"odinger theory.

\subsection{Comparisons}

The two sets of quantum Cauchy surfaces $\hat{\Pi}_t$ and $\hat{\Pi}_\tau$ had each induced a complete set of Dirac observables for $\mathbb H$, along with their conjugate momenta, leading to two distinct Shr\"odinger theories with different sets of fluctuating variables. This is achieved by the exact isomorphisms made possible through \eqref{Dirac observ 2} and \eqref{isomorphism2}.

Observe that the same cannot be achieved by using the quantum relational observables in the form  \eqref{relational observ} in this theory. Using the variable $X_B$ or $X_A$ as the clocks with the specified values, together with $\hat{ O}_{{\bold N}}\equiv (\hat{X}_A,\hat{X}_C,\hat{P}_A,\hat{P}_C)$ or ${}^{*}\!\hat{ O}_{{\bold N}}\equiv (\hat{X}_B,\hat{X}_C,\hat{P}_B,\hat{P}_C)$ the quantum relational observables would be respectively given by
\begin{equation}
\begin{split}
\label{relational observ2}
 \hat{ O}_{{\bold N}} (\bar{X}_B(t))\equiv \int_{-\infty}^{\infty}\,d\lambda \,e^{i \lambda\hat H} \,\,sym\left\{\,\,\hat{ O}_{{\bold N}} \,\,|\hat{P}_B/2m_B|\, \delta(\hat{X}_B-\bar{X}_B(t))\right\}\,\,e^{-i \lambda\hat H}\\ 
 {}^{*}\!\hat{ O}_{{\bold N}} (\bar{X}_A(\tau))\equiv \int_{-\infty}^{\infty}\,d\lambda \,e^{i \lambda\hat H} \,\,sym\left\{\,\,{}^{*}\!\hat{ O}_{{\bold N}} \,\,|\hat{P}_A/2m_A|\, \delta(\hat{X}_A-\bar{X}_A(\tau))\right\}\,\,e^{-i \lambda\hat H}\,. 
\end{split}
\end{equation}
It can be checked that these observables, under generic self-adjoint symmetrizations, do not form an exact representation of $\hat{ O}_{{\bold N}}$ and ${}^{*}\!\hat{ O}_{{\bold N}}$, due to the corrections introduced through the commutations between the clock operators and their conjugate momenta. On the other hand, our observables $\{\big( \hat{\Lambda}_t^{-1} ( \hat{\bold  X}_{{\bold N}}, \hat{\bold  P}_{{\bold N}}) \hat{\Lambda}_t\big)(t)\big\}$ and $\{\big( \hat{{\Lambda}}_\tau^{\!-\!1}({}^{*}\hat{\bold  X}_{{\bold N}}, {}^{*}\hat{\bold  P}_{{\bold N}})\hat{{\Lambda}}_\tau\big)(\tau)\big\}$ constructed as the faithful representations can be shown to take the explicit forms
\begin{eqnarray}
\label{relational observ3}
\big( \hat{\Lambda}_t^{-1} ( \hat{\bold X}_{{\bold N}}, \hat{\bold P}_{{\bold N}}) \hat{\Lambda}_t\big)(t)
&=& \int_{-\infty}^{\infty}\,d\lambda\, \,\,e^{i \lambda\hat H} \,\,\,\,{\sqrt{{|\hat{P}_B^\star|}/2m_B}}\,\,( \hat{\bold  X}_{{\bold N}}, \hat{\bold  P}_{{\bold N}}) \,\,\,\frac{1}{\sqrt{{|\hat{P}_B^\star|}/2m_B}}\nonumber\\
&&\times\sum_{\eta_B}\theta(\eta_B\hat{P}_B)\,\,  \delta(\hat{X}_B-\bar{X}_B(t)) \,|\hat{P}_B/2m_B|\,\theta(\eta_B\hat{P}_B) \,\,\,e^{-i \lambda\hat H}\nonumber\\
\big( \hat{{\Lambda}}_\tau^{\!-\!1}({}^{*}\hat{\bold  X}_{{\bold N}}, {}^{*}\hat{\bold  P}_{{\bold N}})\hat{{\Lambda}}_\tau\big)(\tau) 
&=& \int_{-\infty}^{\infty}\,d\lambda\, \,\,e^{i \lambda\hat H} \,\,\,\,\frac{\sqrt{{|\hat{P}_B|}/2m_B}}{|\phi_{\hat n}(\bar{X}^{+}_{A}(\tau))|}\,\,({}^{*}\hat{\bold  X}_{{\bold N}}, {}^{*}\hat{\bold P}_{{\bold N}}) \,\,\,\frac{|\phi_{\hat n}(\bar{X}^{+}_{A}(\tau))|}{\sqrt{{|\hat{P}_B|}/2m_B}}\nonumber\\
&&\times\theta(\hat{P}_A)\,\, \delta(\hat{X}_A-\bar{X}_A(\tau)) \,|\hat{P}_A/2m_A|\,\,\theta(\hat{P}_A) \,\,\,e^{-i \lambda\hat H}.
\end{eqnarray}
In our observables, the $\hat{\Lambda}_t$ and $\hat{\Lambda}_\tau$ can annihilate with their inverses up to only quantum corrections. Also, we have $\sum_{\eta_B}\theta^2(\eta_B\hat{P}_B)=\hat{I}$ and $\lim_{\hbar\to 0}({}^{*}\hat{\bold  X}_{{\bold N}}, {}^{*}\hat{\bold  P}_{{\bold N}})={}^{*}\!\hat{ O}_{{\bold N}}$. These lead to the fact that 
\begin{equation}
\begin{split}
\label{limits}
\lim_{\hbar \to 0} \big( \hat{\Lambda}_t^{-1} ( \hat{\bold  X}_{{\bold N}},\hat{\bold  P}_{{\bold N}}) \hat{\Lambda}_t\big)(t)=\hat{ O}_{{\bold N}} (\bar{X}_B(t))\,; \,\,\,\lim_{\hbar \to 0} \big( \hat{{\Lambda}}_\tau^{\!-\!1}({}^{*}\hat{\bold  X}_{{\bold N}}, {}^{*}\hat{\bold P}_{{\bold N}})\hat{{\Lambda}}_\tau\big)(\tau) ={}^{*}\!\hat{ O}_{{\bold N}}(\bar{X}_A(\tau))\cdot\theta(\hat{P}_A).
\end{split}
\end{equation}
That is, the two types of observables have the same classical limits in the sectors with a definite sign for the momenta of the clocks. This confirms our expectation from the discussion in the end of Sec.II.A.

Beyond the classical limits, the quantum Cauchy surfaces are fundamental objects in the deep quantum regions, where they define the exact Shr\"odinger theories. Note that while the theory $({\mathbb{D}}, \{\hat{\Pi}_{t}\}_{-\infty\leq t \leq \infty})$ is identical to the corresponding quantum reduced phase space theory, the theory $({}^{*}{\mathbb{D}}, \{\hat{\Pi}_{\tau}\}_{-\infty\leq \tau \leq \infty})$ has no proper correspondence from the reduced phase space method, due to the bounded nature of the classical clock. However, the underlying quantum nature of the clock in $({}^{*}{\mathbb{D}}, \{\hat{\Pi}_{\tau}\}_{-\infty\leq \tau \leq \infty})$ allows the physical time to ``tunnel" to infinity. Thus the theory would always give a purely quantum region for any physical state. Even within the region with the classical limits, we have seen that the observable spectra for $P_B$ and $P_C$ are also corrected by this underlying quantum nature of the clock.

Finally, since both of the Schr\"odinger theories are global ${}^{*}\mathbb{D} = \mathbb D= \mathbb H$ the two theories are dual to each other. The transformation matrix can be calculated through \eqref{trans amp2}, and the result is 
\begin{eqnarray}
\label{trans amp4}
 \big({{}^{*}{\bold P}'_{{\bold N }}}{ ; \, \hat{\Lambda}_{\mathbb S^{\tau}}}\big|{{\bold P}_{{\bold N}}}{ ; \, \hat{\Lambda}_{\mathbb S^{t}}}\big)= e^{-i{{P}_B^\star\,\bar{X}_{B}(t)/\hbar}}\,\, e^{iS(n', \bar{X}_{A}(\tau))}\,\,\delta_{{P}'_B\,,\, {P}_B^\star} \,\delta({P'}_C-{P}_C).
\end{eqnarray}
This represents the change between the two observer frames associated to the two foliations $\{\hat{\Pi}_t\}$ and $\{\hat{\Pi}_\tau\}$.

\section{Summary and Conclusion}

For a Dirac theory of quantum gravity $(\mathbb{H}, \mathbb{K}, \hat{\mathbb{P}})$, we have proposed an exact notion of Cauchy surfaces from the quantum level, which we have argued to be essential for obtaining an effective Schr\"odinger theory. They are generally defined as the right inverse maps of the rigging map $\hat{\mathbb{P}}$.

Similar to its classical counterpart, a quantum Cauchy surface can represent $\mathbb D \subset\mathbb{H}$ with an instantaneous ``quantum reduced phase space" $\mathbb S^{t}\subset \mathbb{K}$. A self-adjoint complete set of operators in $\mathbb S^{t}$ provides a spectrum for the Schr\"odinger wave function describing $\Psi\in \mathbb D$, which is defined at the moment given by the quantum Cauchy surface. Through this representation, a physical fundamental algebra in $\mathbb D$ is also induced by the fundamental algebra in $\mathbb S^{t}$. Further, the quantum degrees of freedom absent in $\mathbb S^{t}$ due to the constraints naturally yield a physical background without any quantum fluctuation, which may provide a notion of time for the wave function. This is very much in analogy to the classical reduced phase space theory. Under a specified background, we also deduced the two essential stabilities for a Schr\"odinger theory to emerge for a finite interval of time. The Heisenberg operators for such a Schr\"odinger theory are labeled by the background field values, and thus they are special Dirac observables in the Dirac theories that are closely related to the standard quantum relational observables.  

We also noted that when the physical domains of two such Schr\"odinger theories overlap, a physical state from the overlapping subspace would have both of the Schr\"odinger representations, related by a unitary transformation between the observer frames associated with the two quantum foliations. Moreover, we argued that each of the Schr\"odinger theories can be written in terms of the relevant path integral transition amplitudes, given by the rigging map matrix elements in the sectors with the specified background. 

According to our formalism, the observer frame of ours can be inferred by the specific background that we observe to be purely classical fields. The significance of such formalism is apparent for deriving a quantum cosmological model from a Dirac theory of quantum gravity. For that, one needs to describe a unitary quantum evolution over a given set of observables, in a fundamentally timeless theory treating all possible quantum fluctuations equally. Furthermore, an emergent Schr\"odinger theory could carry the signatures of the full quantum fluctuations in the underlying Dirac theory, which would be absent in the reduced phase space quantization using the same background. As we have shown through our simple model, this can happen not only to the dynamics, but also to the spectrum of the observables. This special feature revealed by the quantum Cauchy surfaces would be of great interests in the context of quantum cosmology.

For our future works, the most immediate next step should be testing our construction in the Dirac theories of the minisuperspace cosmological models. However, it is more important to implement our proposal in a model theory with the full set of scalar and momentum constraints, so we may show that our formal construction can be realized rigorously. As mentioned, the currently most promising Dirac theory of quantum gravity is loop quantum gravity, which has a solid kinematic Hilbert space of the quantum geometry. In our previous works \cite{CY1}\cite{CY2}, we had applied the Dirac observables of the form \eqref{Dirac observ 1} to a model of loop quantum gravity, to derive the semi classical limits of the model. The model shares same kinematic Hilbert space with the full theory of loop quantum gravity, and is obtained by simplifying the standard scalar constraint operator in the full theory. Although the quantum Cauchy surfaces are applied only in the semi classical limits, the core idea is that our Dirac observables faithfully represent the loop algebra of the quantum geometry in the limits. As a result, the dynamics obtained from using these Dirac observables recovers a specific semi classical limit of general relativity, accompanied by the signature corrections from the quantum geometry. Along this line, our proposal in this paper serves to specify our method right from the quantum level. Therefore correspondingly, it is probably best to improve the previous model so that an implementation of the quantum Cauchy surfaces can be demonstrated in the quantum level. 

There are many other interesting and important topics about quantum gravity we may discuss through the proposal. As what we have argued, the path-integral transition amplitudes may be translated to the components of our Schr\"odinger theories. Further studies in this direction could provide us insights relating the covariant and canonical formulations of quantum gravity, from a more physical point of view. Lastly, we should also address the issue of consistent probabilistic interpretations for the locally defined Schr\"odinger theories suggested by our proposal. 

\section{Acknowledgment}
The author would like to thank Prof. Chopin Soo, Prof. Ho-lai Yu and Prof. Chiang-Mei Chen for the instructive and inspiring discussions. This project is supported in part by the Ministry of Science and Technology of Taiwan under the Grant No. 102-2112-M-008 -015 -MY3.


\begin{thebibliography}{9}
\vspace{0.2em}
\bibitem{kuchartorre}  K. V. Kuchar, \textit{Time and Interpretations of Quantum Gravity}, in the proceedings of
The Fourth Canadian Conference on General Relativity and Relativistic Astrophysics,
edited by G. Kunstatter, D. Vincent, and J. Williams (World Scientific, Singapore
1992).
\vspace{0.2em}
\bibitem{Isham}  C. J. Isham, \textit{Canonical quantum gravity and the problem of time }, gr-qc/9210011, in Integrable Systems, Quantum Groups and Quantum Field Theories,
edited by  L.A. Ibort and M.A. Rodríguez (Kluwer, Dordrecht 1993).
\vspace{0.2em}
\bibitem{kuchar} J. D. Brown, K. V. Kuchar, \textit{Dust as a Standard of Space and Time in Canonical Quantum Gravity}, gr-qc/9409001, Phys. Rev. D51 (1995) 5600
\vspace{0.2em}
\bibitem{torre1} J. D. Romano, C. G. Torre, \textit{Internal Time Formalism for Spacetimes with Two Killing Vectors}, gr-qc/9509055v1, Phys. Rev. D53 (1996) 5634
\vspace{0.2em}
\bibitem{reduced1} K. Giesel, T. Thiemann, \textit{Algebraic Quantum Gravity (AQG) IV. Reduced Phase Space Quantisation of Loop Quantum Gravity}, arXiv:0711.0119, Class. Quant. Grav. 27 (2010) 175009
\vspace{0.2em}
\bibitem{reduced2} W Donnelly, T Jacobson \textit{Hamiltonian structure of Horava gravity}, arXiv:1106.2131, Phys. Rev. D84 (2011) 104019
\vspace{0.2em}
\bibitem{reduced3} C. Soo, H.L. Yu \textit{General Relativity without paradigm of space-time covariance, and resolution of the problem of time},  arXiv:1201.3164, PTEP 2014 (2014) 1, 013E01
\vspace{0.2em}
\bibitem{Dirac1} A. Perez, \textit{Introduction to Loop Quantum Gravity and Spin Foams}, gr-qc/0409061, Lectures given at 2nd International Conference on Fundamental Interactions, Domingos Martins, Espirito Santo, Brazil, 6-12 Jun 2004.
\vspace{0.2em}
\bibitem{Dirac2} D. Giulini, C. Kiefer, \textit{The Canonical Approach to Quantum Gravity: General Ideas and Geometrodynamics}, gr-qc/0611141,  Lect. Notes Phys. 721 (2007) 131
\vspace{0.2em}
\bibitem{ref1} D. Marolf, \textit{Group Averaging and Refined Algebraic Quantization: Where are we now?}, gr-qc/0011112, in Proceedings of the 9th Marcel Grossmann Meeting, edited by V.G. Gurzadyan,
R.T. Jantzen, and R. Ruﬃni (World Scientiﬁc, Singapore, 2002)
\vspace{0.2em}
\bibitem{ref2} D. Giulini, D. Marolf, \textit{On the Generality of Refined Algebraic Quantization},  gr-qc/9812024, Class. Quant. Grav. 16 (1999) 2479
\vspace{0.2em}
\bibitem{ref3} D. Giulini, \textit{Group Averaging and Refined Algebraic Quantization}, gr-qc/0003040v1, Nucl. Phys. Proc. Suppl. 88 (2000) 385
\vspace{0.2em}
\bibitem{lqg1}  A. Ashtekar, \textit{Gravity and the Quantum}, gr-qc/0410054, New J. Phys. 7 (2005) 198
\vspace{0.2em}
\bibitem{lqg2}  C. Rovelli, \textit{Loop Quantum Gravity}, gr-qc/9710008, Living Rev. Rel. (1998) 1:1
\vspace{0.2em}
\bibitem{lqg3}  A. Perez, \textit{Introduction to Loop Quantum Gravity and Spin Foams}, gr-qc/0409061, Lectures given at 2nd International Conference on Fundamental Interactions, Domingos Martins, Espirito Santo, Brazil, 6-12 Jun 2004.
\vspace{0.2em}
\bibitem{reflqc1} K. Noui, A. Perez, K. Vandersloot, \textit{On the Physical Hilbert Space of Loop Quantum Cosmology}, gr-qc/0411039, Phys. Rev. D71 (2005) 044025
\vspace{0.2em}
\bibitem{reflqc2} W. Kaminski, J. Lewandowski, T. Pawlowski, \textit{Quantum constraints, Dirac observables and evolution: group averaging versus Schroedinger picture in LQC
}, arXiv:0907.4322, Class. Quant. Grav. 26 (2009) 245016 
\vspace{0.2em}
\bibitem{reflqc3} W. Kaminski, T. Pawlowski, \textit{The LQC evolution operator of FRW universe with positive cosmological constant
}, arXiv:0912.0162, Phys. Rev. D81 (2010) 024014 
\vspace{0.2em}
\bibitem{reflqc4} D. Marolf, \textit{Quantum Observables and Recollapsing Dynamics
}, gr-qc/9404053, Class. Quant. Grav. 12 (1995) 1199 
\vspace{0.2em}
\bibitem{refqg1} C. Rovelli, \textit{The projector on physical states in loop quantum gravity}, gr-qc/9806121v2, Phys. Rev. D59 (1999) 104015
\vspace{0.2em}
\bibitem{refqg2} D. Marolf, \textit{Group Averaging and Refined Algebraic Quantization: Where are we now?}, gr-qc/0011112v1, in the proceedings of the 9th Marcel Grossmann Conference, Rome 2000
\vspace{0.2em}
\bibitem{anomfree1} M. Assanioussi, J. Lewandowski, I. Mäkinen, \textit{New scalar constraint operator for loop quantum gravity}, arXiv:1506.00299
\vspace{0.2em}
\bibitem{anomfree2} A. Laddha, \textit{Hamiltonian constraint in Euclidean LQG revisited: First hints of off-shell Closure},  arXiv:1401.0931
\vspace{0.2em}
\bibitem{master1} T. Thiemann, \textit{The Phoenix Project: Master Constraint Programme for Loop Quantum Gravity}, gr-qc/0305080, Class. Quant. Grav. 23 (2006) 2211
\vspace{0.2em}
\bibitem{master2} K. Giesel, T. Thiemann, \textit{Algebraic Quantum Gravity (AQG) I. Conceptual Setup}, gr-qc/0607099, Class. Quant. Grav. 24 (2007) 2465
\vspace{0.2em}
\bibitem{master3} B. Dittrich, T. Thiemann, \textit{Testing the Master Constraint Programme for Loop Quantum Gravity III. SL(2,R) Models}, gr-qc/0411140, Class. Quant. Grav. 23 (2006) 1089
\vspace{0.2em}
\bibitem{foam2} C. Rovelli
 \textit{A new look at loop quantum gravity
}, arXiv:1004.1780v4, Class. Quant. Grav. 28 (2011) 114005
\vspace{0.2em}
\bibitem{foam1} J. C. Baez, J. D. Christensen, T. R. Halford, D. C. Tsang,
 \textit{Spin Foam Models of Riemannian Quantum Gravity
}, gr-qc/0202017v4, Class. Quant. Grav.19 (2002) 4627
\vspace{0.2em}
\bibitem{path1} J. B. Hartle, \textit{Spacetime Quantum Mechanics and the Quantum Mechanics of Spacetime
}, gr-qc/9304006, in Gravitation and Quantizations: Proceedings of the 1992 Les Houches Summer School, edited by B. Julia and J. Zinn-Justin (North Holland, Amsterdam, 1995)
\vspace{0.2em}
\bibitem{path2} D. Marolf, \textit{Path Integrals and Instantons in Quantum Gravity
}, gr-qc/9602019, Phys. Rev. D53 (1996) 6979
\vspace{0.2em}
\bibitem{marolf} D. Marolf, \textit{Solving the Problem of Time in Mini-superspace: Measurement of Dirac Observables}, arXiv:0902.1551v1, Phys. Rev. D79 (2009) 084016
\vspace{0.2em}
\bibitem{quantrelat1} S. B. Giddings, D. Marolf, J. B. Hartle, \textit{Observables in effective gravity}, hep-th/0512200, Phys. Rev. D74 (2006) 064018 
\vspace{0.2em}
\bibitem {quantrelat2} C. Rovelli, \textit{Time in quantum gravity: An hypothesis}, Phys. Rev. D43 (1991) 442
\vspace{0.2em}
\bibitem{quantrelat3} D. Marolf, \textit{Quantum Observables and Recollapsing Dynamics}, gr-qc/9404053v5, Class. Quant. Grav. 12 (1995) 1199
\vspace{0.2em}
\bibitem{intro1} A. Ashtekar, \textit{Gravity and the Quantum}, gr-qc/0410054, New J. Phys. 7 (2005) 198
\vspace{0.2em}
\bibitem{intro} C. Rovelli, \textit{Loop Quantum Gravity}, gr-qc/9710008, Living Rev. Rel. (1998) 1:1
\vspace{0.2em}
\bibitem{area} A. Ashtekar, J. Lewandowski, \textit{Quantum Theory of Geometry I: Area Operators}, gr-qc/9602046v2, Class. Quant. Grav. 14 (1997) A55
\vspace{0.2em}
\bibitem{volume} A. Ashtekar, J. Lewandowski, \textit{Quantum Theory of Geometry II: Volume operators}, gr-qc/9711031v1, Adv. Theor. Math. Phys. 1 (1998) 388
\vspace{0.2em}
\bibitem{volume1} T. Thiemann, \textit{Closed Formula for the Matrix Elements of the Volume Operator in Canonical Quantum Gravity}, gr-qc/9606091, J. Math. Phys. 39 (1998) 3347
\vspace{0.2em}
\bibitem{CY1} C.Y. Lin \textit{Emergence of General Relativity from Loop Quantum Gravity: A Summary}, arXiv:1111.2107, Class. Quant. Grav. 29 (2012) 082001
\vspace{0.2em}
\bibitem{CY2} C.Y. Lin \textit{Emergence of Loop Quantum Cosmology from Loop Quantum Gravity: Lowest Order in h},  arXiv:1111.1766
\vspace{0.2em}
\end{thebibliography}
\end{document}